\documentclass[11pt]{article} 
\usepackage[utf8]{inputenc}
\usepackage{multirow}
\usepackage{amsfonts}
\usepackage{amsmath}
\usepackage{amssymb}
\usepackage{amsthm}
\usepackage{mathtools}
\usepackage{dsdshorthand}
\usepackage{tikz}
\usepackage{tensor}
\usetikzlibrary{decorations}
\usetikzlibrary{trees}
\usetikzlibrary{decorations.pathmorphing}
\usetikzlibrary{decorations.markings}
\usetikzlibrary{external}
\usetikzlibrary{intersections}
\usetikzlibrary{shapes,arrows}
\usetikzlibrary{arrows.meta}
\usetikzlibrary{calc}
\usetikzlibrary{shapes.misc}
\usetikzlibrary{decorations.text}
\usetikzlibrary{backgrounds}
\usetikzlibrary{fadings}
\usetikzlibrary{tikzmark,calc,arrows,shapes,decorations.pathreplacing}
\tikzset{
	cross/.style={cross out, draw=black, minimum size=2*(#1-\pgflinewidth), inner sep=0pt, outer sep=0pt},
	branchCut/.style={postaction={decorate},
		snake=zigzag,
		decoration = {snake=zigzag,segment length = 2mm, amplitude = 2mm}	
}}
\usepackage{caption}
\usepackage{color}
\definecolor{darkgreen}{rgb}{0,0.5,0}
\definecolor{darkblue}{rgb}{0,0,0.6}
\definecolor{purple}{rgb}{0.4,.2,0.7}
\usepackage[margin = 2.5cm]{geometry}
\usepackage{graphicx}
\usepackage[hyperindex,breaklinks,hyperfootnotes = false, colorlinks = true, linkcolor = darkblue, citecolor = purple]{hyperref}
\usepackage{subcaption}
\usepackage{ytableau}

\usepackage{comment}

	\newcommand{\ee}{\end{equation}}
\newcommand{\bea}{\begin{eqnarray}}
	\newcommand{\eea}{\end{eqnarray}}
\def\la{\label}
\def\nref#1{(\ref{#1})}
\def\half{{1 \over 2 }}

\begin{document}
	
	\thispagestyle{empty}
	\begin{center}
		~\vspace{5mm}
		
		\vskip 2cm 
		
		{\LARGE \bf 
			Physical instabilities and the phase of the Euclidean path integral 
		}

		\vspace{0.5in}

		Victor Ivo$^1$, Juan Maldacena$^2$, and Zimo Sun$^{1,3}$

		\vspace{0.5in}

		$1$
		{\it  Jadwin Hall, Princeton University,  Princeton, NJ 08540, USA }
		\\
		~
		\\
		$2$
		{\it   Institute for Advanced Study,  Princeton, NJ 08540, USA }
        \\
        ~
		\\
		$3$
        {\it Princeton Gravity Initiative, Princeton University, Princeton, NJ 08544, USA }

	\end{center}
	
	\vspace{0.5in}
	
	\begin{abstract}
	
	We compute the phase of the  Euclidean gravity partition function on manifolds of the form $S^p \times M_q$. We find that the  total phase is equal to the phase in pure gravity on $S^p$ times an extra phase that arises from negative mass squared fields that we obtain when we perform a Kaluza-Klein reduction to $S^p$. The latter can be matched to the phase expected for physical negative modes seen by a static path observer in $dS_p$. In the case of $S^p \times S^q$ the answer can be interpreted in terms of a computation in the static patch of $dS_p$ or $dS_q$. We also provide the phase when we have a product of many spheres. We clarify the procedure for determining the precise phase factor. We discuss some aspects of the interpretation of this phase.

	\end{abstract}
	
	\vspace{1in}
	
	\pagebreak
	
	\setcounter{tocdepth}{3}
	{\hypersetup{linkcolor=black}\tableofcontents}

	\section{Introduction}

	 In this paper,  we focus on the phase of the one loop contribution to the pure gravity path integral  around various solutions of the form $S^p \times M_q$. 
	 We define the phase by performing a small deformation of Newton's constant, or $\hbar $. Euclidean gravity  gives a certain phase that we can write as  	
	  \be \la{AnsPha}
	  Z(S^p \times M_q) = i^{p+2 } (-i)^{\cal N }  \left|Z(S^p \times M_q)\right| ~,~~~~~{\rm for } ~~~~~{ 1 \over G_N} = { 1 \over |G_N|}( 1- i \epsilon) 
	\ee 
	where, for now, this is just the definition of ${\cal N}$. The total phase depends on the number of Euclidean negative modes of certain operators. 
	We show that the number ${\cal N}$ appearing in  \nref{AnsPha} can be viewed as arising from   exponentially growing quasinormal modes in $dS_p \times M_q$. These are modes that, in the static path,   grow as $e^{ \kappa t }$ where $t$ is the proper time.  The precise number ${\cal N}$ depends also on the actual size of $\kappa/H_p$ where $H_p$ is the Hubble constant. 
	 The first factor in \nref{AnsPha}, $i^{p+2}$,  is the same as the factor that Polchinski computed for the sphere case \cite{Polchinski:1988ua, Anninos:2020hfj}.  
	 In \cite{Maldacena:2024spf},  it was suggested that this phase factor could be removed by  including the observer after we focus on the quantity that is supposed to have a counting interpretation. We are saying that the extra phases that arise in \nref{AnsPha} arise from physical instabilities that are manifest as growing modes in the static patch. 
	 
	 As a simple example, to illustrate the idea in this paper, let us go back to the case of an $S^D$ but consider two observers moving along the same trajectory instead of just one observer as in \cite{Maldacena:2024spf}. The second observer introduces an extra phase factor $(-i)^{D-1}$. We can interpret it as arising from true physical unstable modes corresponding to the distance between the two observers. Indeed, if the observers   interact with one another, say they are joined with a spring that is stiff enough to keep them together, then the mode corresponding to the separation would effectively be an upright harmonic oscillator and the extra phase would disappear. On the other hand, if they are non interacting, then the relative distance is an upside down harmonic oscillator, which is the origin of the extra phase factor.
	 
	 In this paper, we discuss the more complicated case where the unstable modes come from the internal manifold. We will always have one unstable mode that comes from the volume of the internal manifold. 
	 
	 We will also discuss special cases where the $M_q$ is a product of spheres and find that   the phase factor has the simple expression 
	\begin{equation} \la{ProdSph}
Z(S^{p_{1}}\times S^{p_{2}} \times .... S^{p_{n}})=(-i)^{(n-2)(D+n+1)}|Z(S^{p_{1}}\times S^{p_{2}} \times .... S^{p_{n}})| ~,~~~~~~{\rm for }~~~ p_i > 2
\end{equation}
where $D=\sum_{i}p_{i}$ and $n$ is the number of spheres, when all $p_{i}>2$. 

In particular, in the case of $n=2$, for $S^p \times S^q$,  we see that there is no phase. This appears to contradict \nref{AnsPha}. However, we find that the number ${\cal N}$ appearing in \nref{AnsPha} depends on which of the spheres we have turned into a de-Sitter space and it is $p$ dependent, even though the total phase in \nref{ProdSph} depends only on the sum of the $p_i$. 

This paper is organized as follows. In section 2, we discuss how Lorentzian unstable modes give rise to factors $(-i)$ in the Euclidean partition function. 
In section 3, we consider a $dS_p \times M_q$ compactification. We discuss the possible negative mass squared fields that can arise through the Kaluza-Klein reduction and compute their contribution to the phase factor. In section 4, we clarify the contour rotation procedure by demanding that the result is independent of gauge choices. We also write the phase of the Euclidean partition function around a general solution. In section 5 we discuss more aspects of the Euclidean partition function evaluating it more directly by using Euclidean methods in manifolds such as $S^p \times M_q$ and $S^p \times S^q$ for general values of $p,~q\geq 2$. We also evaluate the phase for the case of $S^{p_1} \times S^{p_2} \times \cdots \times S^{p_n}$.  In section 6, we discuss some aspects of the case $S^2 \times M_q$, including the connection to two dimensional dilaton gravity and the Hartle-Hawking wave function of the universe. In section 7, we present some conclusions.

 {\bf Note:} As this paper was in preparation,  we learnt that X. Shi and G. Turiaci\cite{Shi:2025amq}, and  S.  Akhtar, R. Mahajan and H. Tang 
 	 were also studying aspects of partition functions on spheres.

	\section{Phase factor from physical negative modes in   theories with no gravity} 
	
	\subsection{ The phase of the Euclidean partition function for an inverted harmonic oscillator } 
	
	Consider an inverted harmonic oscillator with frequency $\omega = -i \kappa$, $\kappa> 0$, with the Euclidean action 
	\be 
	S_E = { 1 \over \hbar } \int_0^\beta dt \half ( \dot x^2 - \kappa^2 x^2 ) ~,~~~~~\kappa > 0 
	\ee 
	Fourier transforming $ x= \sum_n x_n e^{ i  { 2 \pi \over \beta } n t } $ we find negative modes for 
	\be 
	 - n_-  \leq n \leq n_- ~,~~~~~~~ n_- = \left \lfloor {   \kappa \beta \over 2\pi  }  \right \rfloor 
	 \ee 
	 where $\lfloor y  \rfloor$ is the integer part of $y$. 	 Note that $n=0$ is always a negative mode.  Depending on the value of   $\kappa \beta$, there could be additional ones, which come in pairs, $\pm n$. 
	 For simplicity, we assume that $\kappa \beta/(2\pi)$ is not an integer, so that we do not have to worry about zero modes.

	 In order to define the path integral we will continue the overall coefficient of the action in the path integral as 
	 \be 
	 { 1 \over \hbar } \to { 1 \over |\hbar | } (1 - i \epsilon) ~,~~~~~ \epsilon > 0 
	 \ee 
	 Rotating the integral for the negative modes to the imaginary axis, avoiding the direction of maximal increase for the exponent, we get 
	 \be 
	  Z = (-i)^{ 2 n_- +1 } \times ({\rm positive } ) \la{HOPh}
	  \ee 
	  as the phase factor for the path integral. 
	  
	  An important conclusion from this simple example is that a {\it single} harmonic oscillator gives several factors of $(-i)$. The precise number depends on the product $\kappa\beta$.
	  
	  \subsection{ The phase factor for a negative mass squared field   in de Sitter} 
	  \la{SpScalar} 
	  
	  We now consider the case of de-Sitter, $dS_D$,  and a free field with negative mass squared, $m^2 = -   \mu^{2}$, $\mu>0$.   We can compute the phase factor in two different ways that give the same answer. 
	  First, we can consider the static patch, the field can be decomposed into quasinormal modes. 
	  Some of them will grow like $e^{ \kappa t}$.
	  More precisely, the possibly growing quasinormal values of $\kappa$   are given by the formula  
	  \be \la{QNMka}
	  \kappa =\sqrt{ \mu^2 r_D^2 +{ (D-1)^2 \over 4 } }-{(D-1) \over 2}  - n ~,~~~~~n=0,1,2, \cdots 
	  \ee 
	  where $r_D$ is the de-Sitter radius, and $n$ is upper bounded by the condition that $\kappa$ is positive. 
	  
	  For each such mode, we expect to get a factor \nref{HOPh}. 
	  So, from this point of view we get a factor 
	  \be \la{Ndef}
	  (-i)^{\cal N} ~,~~~~~~{\cal N } = \sum_{  \kappa_i} ( 1 + 2 n_-^i) ~,~~~~~~ n_-^i =   \lfloor \kappa_i r_D \rfloor
	  \ee 
	  where the sum is over all the growing quasinormal modes.  
	  	  
	  The other method would be to do the Euclidean path integral over the $D$ dimensional sphere $S^D$. In this case, we can decompose the 
	  field in angular momentum harmonics and then the quadratic term in the action is proportional to  
	  	  \be \la{QuadAc}
	   \phi_{\ell \vec m } \left[  \ell ( \ell + D-1) - \mu^{2} r_D^2 \right] \phi_{\ell,-\vec m } 
	  \ee 
	  where $\vec m$ labels the states within each angular momentum $\ell$ of $SO(D+1)$. 
	  So we have negative modes for 
	  \be \la{Euclll}
	  0 \leq \ell \leq  \ell_- ~,~~~~~~~~~
	  \ee 
	  where $\ell_-$ is the largest integer for which \nref{QuadAc} is negative. 
	  This gives 
	  \be \la{NtildeDef}
	  Z = (-i )^{\cal \widetilde N} ~,~~~~~~ {\cal \widetilde N } = \sum_{\ell=0}^{\ell_-} d^{D+1}_\ell 
	  \ee 
	  where $d_\ell^{D+1}$ is the number of states for a spin $\ell $ representation of $SO(D+1)$ 
	  \be 
	  d^{D+1}_\ell = \left( \begin{array}{c} \ell + D  \\ \ell \end{array} \right) - 	\left( \begin{array}{c} \ell + D-2  \\  \ell -2 \end{array} \right)
	\ee

We will now argue that these two computations give the same result by arguing that ${\cal N} = \widetilde {\cal   N } $. 	  
	 
	 Let us determine the number ${\cal \tilde N}$ as follows. The Lorentzian wave equation can be viewed as  representation of (generically) non integer spin $j$ of $SO(1,D)$ obeying 
	 \be 
	 j ( j + D-1) =\mu^{2}r_{D}^{2} 
	 \ee 
	 The quasinormal mode frequencies correspond to the eigenvalue under an  $SO(1,1)$ subgroup of $SO(1,D)$. 
	  Now we  imagine that we continuously change $\mu$ and we slowly reduce it to zero, and, after that, we make it slightly imaginary so that $m^2 >0$.  
	  As we do this,  all the quasinormal modes that are exponentially increasing will see their imaginary frequencies $\kappa_{  i}$ monotonically reduced, see \nref{QNMka}.   When $\kappa_i r_D $ crosses an  integer we find that ${\cal N}$ is reduced by two, ${\cal N } \to {\cal N} -2$ because $n^i_- \to n^i_- -1$. In addition, when $\kappa_i$ crosses zero, then ${\cal N} \to {\cal N} -1$. Therefore,  we need to count these ``events''. When $\kappa_i r_D $ is an integer or zero, let us call it $\tilde m \in \mathbb{Z}$, then we find a solution of the Euclidean problem on $S^D$.
	  Let us explain this in more detail. The metric in the static patch is 
	  \be  
	  ds^2 =  - \cos^2 \theta dt^2 + d\theta^2 + \sin^2 \theta d\Omega_{D-2}^2 
	  \ee 
	  The Euclidean sphere is obtained by $t \to i \varphi$. So if a mode behaves as $e^{ \tilde m t }$ it becomes a Euclidean mode behaving as $e^{ i \tilde m \varphi}$. So when $\tilde m$ is an integer, we find that this is a well defined mode in the sphere. Also the regularity condition for the quasinormal mode becomes the regularity condition in the Euclidean theory when $\cos^2 \theta \to 0$.  
	  This also means that  
	   $\tilde m$ becomes the  eigenvalue of $SO(2)$ in the $SO(D+1) \to SO(2) \times SO(D-1) $ decomposition of the symmetry group of the Euclidean problem.  This can only happen when $j$ crosses some integer $\ell$ which automatically will be in the range \nref{Euclll}. We can determine how much ${\cal N}$ changes as we cross $j= \ell$ by computing the dimension of the representation, $d^{D+1}_\ell$. The reason is that each state with $\tilde m\not =0$ contributes 2 units to ${\cal N}$, $\pm \tilde m$, and the states with $\tilde m=0$ contribute one unit.  
	 Summing over all the contributions that appear when $j$ crosses all the integer values of $\ell$ in \nref{Euclll} we get the total value of ${\cal N}$, which is then equal to $\tilde { \cal N}$ in 
	 \nref{NtildeDef}.   
	  
	 The conclusion from this discussion is that the phase factors that appear when we consider a negative mass squared field in Euclidean de Sitter, or $S^D$,  can be viewed as arising from each of the physically growing modes that we have in the static patch of Lorentzian de Sitter, or $dS_D$.  We should emphasize that each Lorentzian growing mode gives a number of factors of $(-i)$ that depends on how negative it is, namely $(1+ 2 n^i_-)$ \nref{Ndef}. 
	 Of course, each negative Euclidean mode only gives a single factor of $(-i)$.

	  \section{Instabilities of $dS_p \times M_q$ from Kaluza-Klein theory}
	  \label{dSpMqLorAlt}

In order to find the physical unstable modes of pure gravity on $dS_p \times M_q$ we can use results on the Kaluza-Klein (KK) reduction of the $D$ dimensional gravity theory to $dS_p$. 
This  problem   was analyzed in detail in \cite{Hinterbichler:2013kwa}.  We   focus on the spectrum for solutions of form $dS_{p} \times M_{q}$. 

The action in $D$ dimensions is 
\begin{equation}
I=\frac{1}{16 \pi G_{N}}\int (R-2 \Lambda)
\end{equation}
with $\Lambda>0$ the cosmological constant. Einstein equations then fix the Ricci tensor in the manifold to be
\begin{equation}
R_{a b}={(D-1) \over r_D^2 } g_{a b} \text{  , ~~~~~ with  ~~ } \Lambda=\frac{(D-1)(D-2)}{2 r_D^2}   ~,~~~~~~~~r_D=1
\end{equation}
where to simplify the equations we set the curvature radius $r_{D}$ of the manifold to one. Einstein equations thus fix the Ricci scalars $R_{p}$ and $R_{q}$ of $dS_{p}$ and $M_{q}$ respectively, and their associated curvature radii $r^{2}$, to 
\begin{equation}
R_{p}=\frac{p(p-1)}{r_{p}^{2}} ~~\rightarrow ~~r_{p}^{2}=\frac{(p-1)}{(D-1)} \text{ , ~~~~and ~~~} r_{q}^{2}=\frac{(q-1)}{(D-1)}
\end{equation}
where $D= p+q$. 

It is also useful to define the notation we use for the Laplacians restricted to $dS_{p}$ and $M_{q}$ 
\begin{equation}
\nabla_{p}^{2}=g^{\alpha \beta}\nabla_{\alpha}\nabla_{\beta} \text{ , } ~~~~~~~~~~ \nabla_{q}^{2}=g^{\alpha' \beta'}\nabla_{\alpha'}\nabla_{\beta'}
\end{equation}
with $\alpha,\beta$ indices in the $dS_{p}$ factor, $\alpha',\beta'$ indices on the $M_{q}$ factor, and $\nabla_{a}$ the usual covariant derivative in the full manifold $dS_{p}´\times M_{q}$. Having setup this background, we just need to focus on the possible negative mass squared fields that we can obtain in the Kaluza-Klein theory, which will give rise to instabilities, as discussed in section \ref{SpScalar}.

The modes of $D$ dimensional gravity can give rise to spin two, spin one and spin zero fields from the $dS_p$ point of view. It turns out that the spin two or spin one fields only give rise to massless or massive modes. Of course, the massless modes correspond to gauge symmetries. An interesting observation is that the massive spin two modes always obey the Higuchi bound \cite{Higuchi:1986py} thanks to the properties of the spectrum of the Laplacian \cite{Hinterbichler:2013kwa}.
So the conclusion is that we only need to focus  on modes that are scalars on $dS_p$. 
The possible   negative scalar modes are described in section 5.6 of \cite{Hinterbichler:2013kwa}.
 These modes are divided into, in (almost) their notation, as $\phi_{0}$, $F_{n}$, and $\phi^{I}$.  
Let us describe each of them in turn. 

{\bf The volume modulus, $\phi_0$: }

 This is a field that is constant on $M_q$ and parametrizes 
 metric fluctuations of the form,  for  $p>2$, 
\begin{equation} \la{VolModF}
	h_{\alpha \beta}=-\frac{g_{\alpha \beta}}{p-2}\phi_{0} \text{  , ~~~~~~~~ } h_{\alpha' \beta'}=\frac{g_{\alpha' \beta'}}{q}\phi_{0} ~,~~~~~~~{\rm for } ~~~~p>2
\end{equation}

These are fluctuations of the volume of the internal manifold $M_{q}$ which keep the Einstein frame metric fixed in $dS_{p}$. They have a negative mass squared   
\begin{equation}
	r_p^2 m_{0}^{2}=-2(p-1). \la{MassVolM}
\end{equation} 
in units of the $p$ dimensional de Sitter radius, $r_p$ (for an intuition where the negative mass comes from see \nref{potphi0}). These are modes in de Sitter whose negative mass squared is big enough that their $\ell=0$ and $\ell=1$ modes in Euclidean de Sitter are negative modes. In total we have   $1+p+1=p+2$ negative modes,   
or factors of $(-i)$. 

For $p=2$ the volume modulus is not a propagating field, $\phi_0$  behaves as the dilaton of a dilaton gravity in $S^{2}$  \cite{Hinterbichler:2013kwa}. We discuss this case in section \ref{TwoMq}.

{\bf  Higher modes of the scalar Laplacian in the internal manifold, $F_n$: }

The $F_{n}$ modes, unlike $\phi_{0}$, do depend on $M_{q}$, but the specific form of their metric is gauge dependent and not very illuminating.  The relevant metric is composed of terms proportional to $g_{\alpha \beta}$ and $g_{\alpha' \beta'}$ in each of the factors, and also double derivatives of the  scalar mode. The scalars $F_{n}$ correspond to eigenmodes of the scalar Laplacian on $M_q$,  $-\nabla_{q}^{2}$,  with eigenvalue above the conformal Killing vector value \nref{boundnabp}. Their mass squared in units of de Sitter radius is given by

\begin{equation} \la{MassSq}
	r_p^2 m_{n}^{2} = (p-1) \left( -2 + {\hat \lambda_{n} \over q-1 } \right)  
\end{equation}
\be \la{lowBou}
{\rm where }~~~~-  r_q^2 \nabla_q^2 F_n = \hat \lambda_n F_n ~,~~~~~~~~  q <  \hat \lambda_n 
\ee 

Note that the non-zero spectrum of the Laplacian starts at least at $\hat \lambda_{n}= q$, see appendix \nref{ScalarLap},  and the modes that are exactly at this value are removed. The Laplacian achieves that value only for scalars whose gradients are conformal Killing vectors, a case that can happen only when $M_q=S^q$ \cite{Obata1962CertainCF}. Of course,  we are interested only in the values of $\hat \lambda_n$ where \nref{MassSq} is negative. 
Due to the lower bound in \nref{lowBou}, 
the mass squared in \nref{MassSq} is such that for every negative value of $m_{n}^{2}$ only their mode with $\ell_{p}=0$  leads to a negative mode in the Euclidean problem (see \nref{Delta0doub}). 
Their total number of negative modes of the Euclidean problem is the number of eigenvalues of \nref{lowBou} such that $\hat \lambda_n < 2(q-1)$. 

 We call the number of such modes ${\cal N}_{L,q}$, the $L$ index standing for longitudinal since they always have longitudinal fluctuations unlike $\phi_{0}$ and $\phi^{I}$.  We can also view them as higher Kaluza-Klein modes of the volume field. 

{\bf  Eigenfunctions of the tensor Laplacian in the internal manifold, $\phi^I$: }

The remaining possible negative modes are the scalars $\phi^{I}$ which correspond to metric deformations in $M_{q}$ which are transverse and traceless, that is
\begin{equation}
	h_{\alpha' \beta'}=\phi^{I}\psi_{\alpha'\beta',I}
\end{equation}
where $\psi_{\alpha'\beta',I}$ are tensor eigenmodes in $M_{q}$ satisfying 
\begin{equation} \la{TTEigeV}
	[-\nabla_{q}^{2}\psi_{\alpha'\beta',I}-2\tensor{R}{_{\alpha'}^{\gamma'}_{\beta'}^{\delta'}}\psi_{\gamma'\delta',I}]= \lambda_{I} \psi_{\alpha'\beta',I} \text{, } ~~~~~~~ \nabla^{\beta'}\psi_{\alpha'\beta',I}=g^{\alpha'\beta'}\psi_{\alpha'\beta',I}=0
\end{equation} 
and with $I$ being just an index labelling the eigenmode. These modes correspond to negative mass squared scalars in de Sitter with   
\begin{equation}
	r_{p}^{2}m_{I}^{2}=\frac{(p-1)}{(D-1)}\lambda_{I}
\end{equation}
in units of the $dS_{p}$ radius $r_{p}$ \footnote{Note that our definition for $\lambda_{I}$ is slightly different from that in \cite{Hinterbichler:2013kwa}, since there they define $\lambda_{I}$   as an eigenvalue of the Lichnerowickz operator in $M_{q}$, which is slightly different from the differential operator we wrote in \nref{TTEigeV}. However, the operators  only differ by a constant shift, and the final masses of the fields on $S^p$ are the same.  }.
Each of these fields  contribute  some factors of $(-i)$, as in   \nref{NtildeDef}, to the Euclidean  path integral. 
 Note that each field can give rise to many negative modes if its mass $m^2_I$ is sufficiently negative. However, since they are just scalar fields in de Sitter, the QFT argument discussed in subsection \nref{SpScalar} implies that this number of Euclidean negative modes \nref{NtildeDef}   matches the number of negative modes coming from the weighted quasinormal mode counting \nref{Ndef}.
 
 Adding up the contribution from all the scalar fields that come from the    transverse traceless tensors in $M_{q}$ gives us a total number of negative modes that we  denote by   ${\cal N}_{TT,q}$. This means that their total contribution to the phase is $(-i)^{{\cal N}_{TT,q}}$. 

Putting all these results together, the number of physical negative modes   is 
\be \la{CalNd}
{\cal N} = p+2+{\cal N}_{L,q}+{\cal N}_{TT,q}
\ee 
 By ``physical'' we mean that these negative modes can also be interpreted as coming from instabilities of the Lorentzian theory as discussed in section \ref{SpScalar}. 

We can now compute the phase of the path integral around  
$S^{p} \times M_{q}$ by using these Kaluza-Klein results. In other words, the path integral can be written as the  gravity path integral on  $S^{p}$ times the path integral over all the Kaluza-Klein modes, each of which is some field on $S^p$. 
Each of the negative mass squared scalars we discussed above will contribute some factors of $(-i)$ to the path integral. Then the total phase of the path integral is, for $p>2$, 
\begin{equation}
\label{phaseKK}
Z[S^{p}\times M_{q}]=Z[S^{p}]\times Z_{KK}=i^{p+2}(-i)^{p+2+{\cal N}_{L,q}+{\cal N}_{TT,q}}(\text{positive})=(-i)^{{\cal N}_{L,q}+{\cal N}_{TT,q}}(\text{positive})
\end{equation}

In particular, the expression \nref{phaseKK} makes manifest that $\cal{N}$ as defined in \nref{AnsPha} and \nref{CalNd} is always captured by the phase of $Z_{K K}$, which is precisely the number of physical negative modes in $dS_{p} \times M_{q}$. The mismatch of the phase of the partition function  and $(-i)^{\cal{N}}$ is therefore precisely the phase of the sphere partition function in $S^{p}$.

In section \ref{EucPI}, we check  that we obtain the same phase for the path integral from pure gravity in the full $D$ dimensional perspective, and we evaluate it explicitly for a couple of examples.    
	  
	  \section{Understanding the procedure for rotating the scale factor } 
	  	\label{ContFix}
	  
	  In order to define the phase of the path integral we have chosen the continuation 
	  \be \la{GNCont}
	  { 1 \over G_N }  = { 1 \over |G_N|} (1 - i \epsilon ) ~,~~~~~~ 1\gg \epsilon > 0 
	  \ee 
	  In principle,  we could choose a different sign for $\epsilon$ and we would need to replace all $i \to -i$ below. But we choose this to be definite. 
	  
	  As explained in \cite{Polchinski:1988ua}, the phase of the partition function on $S^D$ arises from the rotation of the contour of the scale factor. But the precise sign of the $i$ was not determined. 
	  Of course, we expect that the sign of the $i$ depends on the choice \nref{GNCont}. But even after this choice there are two possible procedures we could follow to determine the factor of $i$ that give two different answers. Here we will explain what the right procedure is.

	  \def\betagauge{\eta} 
	  
	  We first discuss the setup. We start with the $D$ dimensional Euclidean action for Einstein gravity with a positive cosmological constant
	  \be 
	  I_E = - { 1 \over 16 \pi G_N } \int \sqrt{g} ( R - 2 \Lambda ) ~,~~~~~~ 2\Lambda = (D-1) (D-2) ~,~~~~~~~~r_D=1
	  \ee 
	  where we have set the length scale so that the sphere solution has radius one, $r_D=1$. Of course,  we could also consider other solutions of 
	  the equations of motion 
	  \be \la{EinstCond}
	  R_{a b } = { 1 \over D} g_{a b } R = (D-1) g_{a b } 
	  \ee 
	  which are not  spheres. 
	  
	  We consider a solution and expand the metric around the solution as 
	  \be 
	  g_{a b } \to  \hat g_{a b } =   g_{a b } + h_{a b } 
	  \ee 
	  
	  The quadratic action for $h_{a b}$ can be written, after using the background equations of motion, in the compact form  
	  \be \la{GenQuad}
	  I_{E}=\frac{1}{64 \pi G_{N}}\int \tilde{h}^{a b}(-\nabla^{2}h_{a b}+2 \nabla_{(a}\nabla^{c}\tilde{h}_{b)c}-2\tensor{R}{_a^c_b^d}h_{c d})
	  \ee
	  where
	  \be \la{Tildeh}
	  \tilde{h}_{a b}=h_{a b}-\frac{1}{2}g_{a b}h  ~,~~~~ h=  h^a_a 
	  \ee
	  We decompose $h_{a b}$   into transverse traceless parts, a pure trace part, and a gauge transformation as          \be
	  \label{decomp1}
	  h_{a b}=\psi_{a b}+g_{a b}\phi+2 \nabla_{(a}\xi_{b)} ~,~~~~~~~~{\rm with} ~~ \nabla_a \psi^a_{b} =0 ~,~~~~\psi^a_a =0 
	  \ee
	  where to make the decomposition unique we require $\xi^{a}$ to be orthogonal to Killing vectors and conformal Killing vectors, similarly as in \cite{Law:2020cpj}\footnote{ One might worry that this decomposition is not orthogonal, but one could imagine we do the decomposition as  
	  	  	\begin{equation}
	  		h_{a b}=\psi_{a b}+g_{a b}\hat{\phi}+2 \nabla_{(a}\xi_{b)}- 2\frac{g_{a b}\nabla^{c}\xi_{c}}{D}
	  	\end{equation}
	  	with $\hat{\phi}=\phi+\frac{2 \nabla_{c}\xi^{c}}{D}$. In this decomposition all three pieces are orthogonal to each other and we can derive a measure for the path integral in terms of $\hat{\phi}$, $\psi_{a b}$ and $\xi_{a}$. Then, we can do a redefinition of the field $\hat{\phi}=\phi+\frac{2 \nabla_{c}\xi^{c}}{D}$ which has a trivial Jacobian, and we go back to the decomposition \nref{decomp1}.}.
	  
	  Using \nref{decomp1} the action becomes 
	  \be \la{ActDec}
	  I_{E}=\frac{1}{64 \pi G_{N}}\int [\psi^{a b}(-\nabla^{2}\psi_{ab}-2\tensor{R}{_a^c_b^d}\psi_{c d})-(D-1)(D-2)\phi(-\nabla^{2}-D)\phi]
	  \ee 
	  As expected, we see that the gauge modes $\xi^a $ disappeared from this action. 
	  
	 For convenience,  we can add a gauge fixing term to the action of the form
	  \begin{equation}
	  \label{gaugefix}
	  	I_{gf}=\frac{\alpha}{32 \pi G_{N}}\int \bigg[\nabla^{b}h_{a b}-\betagauge \nabla_{a}h\bigg]\bigg[\nabla_{c}h^{a c}-\betagauge \nabla^{a}h\bigg]
	  \end{equation}
	  
	  In the de Donder gauge $(\betagauge =\half $) with $\alpha =1$ we get an action for the scalar of the form 
	  \be \la{MLa}
	  \exp\left\{  + c (1-i \epsilon) \int h [ - \nabla^2 - 2 (D-1) ] h \right\} ~,~~~~~~c> 0 
	  \ee 
	  
	  We now will explore two procedures. 
	  
	  \underline{\bf  Procedure I:} 
	  
	  We start with the integral \nref{MLa} defined over real $h$ and we rotate the contour only for the modes that have non-convergent Gaussians. We decide the direction of rotation by demanding that we do not cross the direction of maximal increase which is along $|h|e^{ i\epsilon/2}$. This means that we rotate as
	  \be    
	  h = - i \gamma 
	  \ee 
	  and integrate over real $\gamma $ for such modes. 
	  This then gives a factor 
	  \be \la{ZIphase}
	  Z^I_{\rm phase} = (-i)^{\infty -(D+2) } = i^{D+2} 
	  \ee 
	  where the $D+2$ comes from the modes we do not rotate (the $\ell= 0,1$ modes). We have also assumed that $(-i)^{\infty}$ is an ultralocal term that we can drop. 
	  
	  \underline{\bf  Procedure II:} 
	  
	  Here we first rotate the whole field to $h \to \pm i \gamma$ and we discard the $(\pm i)^\infty $ factor. After that,  we get an  integrand of the form  
	  \be \la{MLaga}
	  \exp\left\{  - c (1-i \epsilon) \int \gamma [ - \nabla^2 - 2 (D-1) ] \gamma \right\} ~,~~~~~~c> 0 
	  \ee 
	  Since most of the modes are now convergent, one might be tempted to argue that this is the more fundamental definition of the path integral. 
	  We now rotate only the $D+2$ negative modes of this action, again avoiding the direction of maximal increase which is along $\gamma = |\gamma|   e^{ i \epsilon/2}$. This means that we rotate as 
	  \be 
	  \gamma = - i h 
	  \ee 
	  and integrate over real $h$ for those modes. 
	  This gives 
	  \be 
	  Z^{II}_{\rm phase} = (-i)^{D+2} 
	  \ee 
	  which does not agree with \nref{ZIphase} when $D$ is odd. 
	  
	  This second procedure was followed in version one of \cite{Maldacena:2024spf} (but with a different sign for $\epsilon$). 
	  
	  On other manifolds, the choice between procedure I or II would also give different answers.

	  \subsubsection{Determining the right procedure }

	 In order to determine the right procedure we will compute the determinant using a more general gauge choice, for general $\alpha$ and $\betagauge$ in \nref{gaugefix} and demand that the answer is gauge invariant. Spoiler alert, the right answer will be procedure I.

	 For a more general gauge choice the action is more complicated, and in order to disentangle the many terms appearing in the action it is useful to decompose  $h_{a b}$ in \nref{decomp1} further as
	  \begin{equation} \la{EqnDec}
	  	h_{a b}=\psi_{a b}+g_{a b}\phi+2 \nabla_{(a}\xi_{b)}^{T}+\nabla_{a}\nabla_{b}\chi
	  \end{equation}
	  with $\xi_{a}^{T}$ a transverse vector.  To avoid redundancy we require $\xi_{a}^{T}$ to be orthogonal to isometries and $\nabla_{a}\chi$ to be orthogonal to conformal Killing vectors, which are derivatives of the $\ell=1$ modes in the sphere. This means that the field $\chi $ is missing the $\ell =0,1$ modes.  Since the last two terms in \nref{EqnDec} are gauge transformations, the Einstein Hilbert action does not depend on them. In fact, for the $S^D$ case,  
	  the full action  for these fluctuations is  
	  \begin{equation}
	  	\begin{gathered}
	  		I_{E}+I_{gf}=\frac{1}{64 \pi G_{N}}\int \bigg\{\psi^{a b}(-\nabla^{2}+2)\psi_{a b}+\frac{\tilde{\alpha}(D-1)(D-2)}{(D\betagauge-1)^{2}} \xi_{a}^{T}\left[-\nabla^{2}-(D-1)\right]^{2}\xi_{a}^{T}\\
	  		-(D-1)(D-2)\tilde{\phi}\left[ -(1-\tilde{\alpha})\nabla^{2}-D\right]\tilde{\phi}\\
	  		+\frac{\tilde{\alpha}(D-1)(D-2)}{ (D\betagauge-1)^{2}} \chi \left[  (-\nabla^{2})\frac{(-\nabla^{2}-D)[(1-\betagauge)(-\nabla^{2})-(D-1)]^{2} }{[(1-\tilde{\alpha})(-\nabla^{2})-D]} \right] \chi \bigg\}
	  	\end{gathered}
	  \end{equation}
	  with
	  \begin{equation} \la{tiPhiDe}
	  	\phi=\tilde{\phi}+\frac{\tilde{\alpha}(-\nabla^{2})[(1-\betagauge)(-\nabla^{2})-(D-1)]}{(D\betagauge-1)[(1-\tilde{\alpha})(-\nabla^{2})-D]}\chi ~,~~~~~~~~~\tilde{\alpha}\equiv \frac{2\alpha(D\betagauge-1)^{2}}{(D-1)(D-2)}
	  \end{equation}
	  We have decoupled the original variables $\phi$ and $\chi$ by defining the new variable $\tilde \phi$  in \nref{tiPhiDe}.  Note that, the Jacobian for the  change of  variables from $\phi$ to $\tilde{\phi}$ is trivial. We then could imagine that we do the integral over $\tilde{\phi}$ first, in such a way that, after the displacement by $\chi$, the contour for $\tilde{\phi}$ is the same as the contour for $\phi$. 
	  
	  The first eigenmodes of $\tilde{\phi}$ with $\ell=0$ and $\ell=1$ are both positive as long as $\tilde{\alpha}>0$\footnote{For the discussion of the $\tilde{\alpha}<0$ case,  see Appendix \nref{moreon}}. The same modes in $\chi$ are excluded since $\ell=0$ is a constant whose contribution to \nref{EqnDec} vanishes and $\ell=1$ corresponds to modes whose derivatives are conformal Killing vectors. The next modes for $\chi$ will have the same sign as the sign of $G_{\alpha}=(1-\tilde{\alpha})(-\nabla^{2})-D$, since all the other operators in its action are positive for $\ell>1$. Therefore, for $\ell>1$ we have that the action for $\tilde{\phi}$ and $\chi$ will appear with opposite signs. A positive mode for $G_{\alpha}$ will define a negative mode for $\tilde{\phi}$ and a positive mode for $\chi$, and the converse is true as well.
	  
	  If $1>\tilde{\alpha}>0$,  the spectrum of $G_{\alpha}$ is completely positive above some maximum value of angular momentum $\ell_{max}$, meaning that $\tilde{\phi}$ has only negative eigenvalues above this value and $\chi$ only positive eigenvalues above it. However, there are a finite number of modes from $0$ to $\ell_{max}$ that have positive action for $\tilde{\phi}$. We denote this total number by $n_{E}$. These modes also have a    negative action for $\chi$, except for $\ell=0$ and $\ell=1$ modes, which are removed from the $\chi$ path integral. It is important to note that $n_{E}$ depends explicitly on the gauge fixing parameter $\alpha$, in particular by making $\tilde{\alpha} \rightarrow 1$ from below we can make $\ell_{max}$ as big as we want. 
	  
	  We now discuss the evaluation of the phase of the path integral with this gauge fixing procedure, applying  procedure I and   procedure II. 
	  
	  According to procedure I, we should rotate the integration contour of the $\tilde{\phi}$ modes that are negative, but keep the ones which are zero or positive along the real axis. This gives an overall phase of $(-i)^{\infty-n_{E}}=i^{n_{E}}$, since we can drop the ultralocal factor of $i^{\infty}$. We also have to rotate the negative modes of $\chi$, but the $\ell=0$ and $\ell=1$ modes are excluded from $\chi$, such that it only has $n_{E}-(1+D+1)=n_{E}-(D+2)$ negative modes. Rotating the contour for these $\chi$ modes gives a phase of $(-i)^{n_{E}-(D+2)}$. Therefore, according to procedure I the phase of the path integral is $i^{n_{E}-n_{E}+(D+2)}=i^{D+2}$, which is independent of the parameters in the gauge fixing procedure,  and matches Polchinski's original result \cite{Polchinski:1988ua}. 
	  
	  However, according to procedure 2 the contour of $\tilde{\phi}$ starts along the imaginary axis, and rotating the zero and positive modes of $\tilde{\phi}$ back to the real axis would give a phase of $(-i)^{n_{E}}$ instead. The phase coming from $\chi$ would be the same as in procedure I, so the overall phase of the path integral would be $i^{D+2}(-i)^{2 n_{E}}$, which is manifestly gauge dependent because $n_{E}$ is (it depends on the parameters of the gauge fixing procedure). Therefore, procedure II gives an answer for the phase which is not gauge invariant, and therefore is not appropriate.
	  
	  The spectrum changes perhaps the most when $\tilde{\alpha}>1$, in which case the spectrum of $\chi$ is the one which is mostly negative, while the spectrum of $\tilde \phi$ is purely positive. With the procedure I we would rotate all the modes of $\chi$ except $\ell=0$ and $\ell=1$ which are excluded, which gives a phase of $(-i)^{\infty-(D+2)}=i^{D+2}$ for the path integral, which is consistent with the other values of $\tilde{\alpha}$. 
	  
	  The conclusion is that procedure I is the correct one. One might wonder if this conclusion persists if we study the number of negative modes using a more conventional eigenvalue problem. We discuss this and other details, such as what happens for negative gauge fixing terms, in Appendix \nref{moreon}.
	    
	  \subsection{The phase for a general manifold}
	  \la{PhaseGen}
	  
	  Instead of adding the gauge fixing term, we will follow a different procedure for evaluating the phase of the partition function. 
We write the metric as in \nref{decomp1} and then derive an associated measure 
	  for the fields in the path integral, this was the technique used in \cite{Law:2020cpj}. The integral over the gauge transformations $\xi_{a}$ then cancels most of the volume of the diffeomorphism group, except for the part coming from isometries. We are left with a Jacobian from the change of variables involving the operator acting on $\xi$ in \nref{decomp1}, but that is positive. 
	  
	  Let us discuss the negative modes of the action \nref{ActDec}. The transverse traceless tensors $\psi_{a b}$ could have some number of negative modes, which we denote by ${\cal N}_{T T}$. 
	  Then we turn to the action for $\phi$. Let us recall some properties of the spectrum of the operator   $-\nabla^{2}$ on a compact Euclidean Einstein manifold, see appendix \ref{CompEin}.  It has one zero mode corresponding to the constant function, then there is a gap to the first excited mode that is at least $D$. The gap is only equal to $D$ if the manifold has conformal Killing vectors, and these eigenmodes are given by the divergence of conformal Killing vectors. All the other eigenvalues are bigger than $D$, which means that we have an infinite number of negative modes for $\phi$.   The only exceptions are the constant mode which is a good positive mode of the action and the conformal Killing vector ones which should not be included in the $\phi$ path integral because we can now choose instead to include them in the gauge transformations in \nref{decomp1}.  More explicitly, these are scalar modes of the form  
	  \be 
	  \nabla \cdot \xi^{CKV} ~,~~~~~{\rm with }~~~ \nabla_a \xi_{b}^{CKV} + \nabla_b \xi_{a}^{ CKV} = { 2 \over D } g_{ab} \nabla \cdot \xi^{CKV} \not =0  \la{CKVCo}
	  \ee  
	Of course, since they are reparameterizations, they are zero modes of the Einstein action. Let us denote by  ${\cal N}_{CKV}$ the number of such modes in the manifold.
	  
	  We now rotate the negative modes of $\phi$ while not rotating the positive and zero modes, which is the rotation procedure we justified in the beginning of this section. If we rotated the full $\phi$ field,  it would give no phase. It would just be an ultralocal term that should be absorbed in a redefinition of the local counter terms.   However, there are two sets of modes that we are not rotating, first the constant mode because it was originally a positive mode.  If the manifold has conformal Killing vectors, then we are not rotating the scalar modes whose derivatives are conformal Killing vectors \nref{CKVCo}   because we are not integrating over these modes. 
	  The final phase factor from the path integral is then
	  \begin{equation}
	  	\prod_{\text{neg modes}} (-i)=   (-i)^{-1-{\cal N}_{CKV}}\prod_{\rm all ~modes}(-i)=i^{1+{\cal N}_{CKV}},	
	  \end{equation}
	  Putting it all together we get 
	  \begin{equation}
	  	\label{phaseM}
	  	Z(M)=i ^{ 1 +{\cal N}_{CKV} }  (-i)^{{\cal N}_{TT} }|Z(M)|.	
	  \end{equation} 
	  For the sphere, $M=S^D$,  we have that ${\cal N}_{TT} =0$, since the Riemann tensor term in \nref{ActDec} becomes just a positive contribution, and ${\cal N}_{CKV}=D+1$. Then   we recover the Polchinski's result  \cite{Polchinski:1988ua}
	  \be 
	  Z(S^D) = i^{ D+2} |Z(S^D) | 
	  \ee 
	  
	  Another interesting point is that the sphere is the only Einstein manifold of constant curvature that has conformal Killing vectors \cite{Obata1962CertainCF}. So, if we are not discussing the sphere, ${\cal N}_{CKV}=0$ and the $\phi$ integral only contributes a single factor of $i$.   Therefore, to compute the phase of the path integral we just need to find the number of negative modes of the transverse traceless action in the manifold
	  \begin{equation}
	  	\label{TTact}
	  	I_{TT}=\frac{1}{64 \pi G_{N}}\int [\psi^{a b}(-\nabla^{2}\psi_{ab}-2 \tensor{R}{_a^c_b^d}\psi_{cd})]
	  \end{equation}
	  
	  We use this result in section \nref{EucPI} to show that the phase of the path integral in $S^{p} \times M_{q}$ matches the KK result \nref{phaseKK}, and to compute the phase explicitly in a couple of examples.
	  
	 \section{The phase in the Euclidean partition function for  gravity on various manifolds }
	 \label{EucPI}

      \subsection{$S^{p} \times M_{q}$}
Using the decomposition discussed in \nref{decomp1} we can find the phase of the path integral \nref{phaseM} in a generic manifold by analyzing the number of negative modes which are transverse and traceless in the full manifold. Here we find a general strategy to determine them when the manifold is $S^{p} \times M_{q}$.

The first motivation for doing that is to have a computation of the phase that is more closely related to a more usual $D$ dimensional pure gravity picture for the path integral, where we understand better the appropriate contour prescriptions to obtain phases. The second motivation is to discuss the necessary tools to find the phase in specific manifolds. The third motivation is that we can also treat the $p=2$ case.

Equation \nref{phaseM} implies that to know the phase of the partition function we just need to find the number of transverse traceless negative modes in the manifold, since ${\cal N}_{CKV}=0$ for every manifold that is not the sphere. The strategy to compute the number of such modes is to apply a decomposition of metric into transverse and longitudinal parts at each factor of the product, and to the off diagonal components as well. We use indices $\alpha,\beta$ for $S^{p}$ and $\alpha',\beta'$ for $M_{q}$. By doing so we can decompose a general traceless metric fluctuation in $S^{p}\times M_{q}$ as
\begin{equation}
	\begin{gathered}
		\label{decompprod}
		h_{\alpha \beta}=g_{\alpha \beta }\frac{\sigma}{p}+2\nabla_{(\alpha}x_{\beta)}+\nabla_{\alpha}\nabla_{\beta}\chi-\frac{g_{\alpha \beta}}{p}\nabla_{p}^{2}\chi+\psi_{\alpha \beta}\\
		h_{\alpha \beta'}=2 \nabla_{(\alpha}z_{\beta')}+\nabla_{\alpha}\nabla_{\beta'}\zeta + \psi_{\alpha \beta'} \\
		h_{\alpha' \beta'}=-\frac{\sigma}{q}g_{\alpha' \beta'}+2\nabla_{(\alpha'}y_{\beta')}+\nabla_{\alpha'}\nabla_{\beta'}\gamma-\frac{g_{\alpha' \beta'}}{q}\nabla_{q}^{2}\gamma +\psi_{\alpha'\beta'}
	\end{gathered}
\end{equation}
where the $x$, $y$ and $z$ vectors are all transverse in their own factors, so that $\nabla^{\alpha}x_{\alpha}=\nabla^{\alpha'}y_{\alpha'}=\nabla^{\alpha}z_{\alpha}=\nabla^{\alpha'}z_{\alpha'}=0$, and so are the tensors $\psi$, $\nabla^{\alpha}\psi_{\alpha \beta}=\nabla^{\alpha'}\psi_{\alpha'\beta'}=\nabla^{\alpha}\psi_{\alpha\beta'}=\nabla^{\beta'}\psi_{\alpha \beta'}=0$, on top of being traceless,   $g^{\alpha \beta}\psi_{\alpha \beta}=g^{\alpha' \beta'}\psi_{\alpha' \beta'}=0$. 

We also drop the modes of $\chi$ and $\gamma$ corresponding to constants or modes whose gradients are conformal Killing vectors, and the modes of $\zeta$ that are constant in one of the manifolds, since they do not contribute to the metric. We also exclude the combinations of $x_{\alpha}$, $y_{\alpha}$, $z_{\alpha}$ and $z_{\alpha'}$ that correspond to isometries for the same reason. To find the transverse traceless modes, we only need to then finally impose that the metric in \nref{decompprod} is transverse in the full manifold, which leads to (using formulas in appendix \ref{CompEin}) 
\begin{equation}
	\begin{gathered} \la{TosC}
		[\nabla_{p}^{2}+(D-1)]x_{\alpha}+\nabla_{q}^{2}z_{\alpha}=[\nabla_{q}^{2}+(D-1)]y_{\alpha'}+\nabla_{p}^{2}z_{\alpha'}=0\\
		\nabla_{\alpha}\bigg[\frac{\sigma}{p}+\frac{(p-1)}{p}\nabla_{p}^{2}\chi+(D-1)\chi+\nabla_{q}^{2}\zeta\bigg]=0\\
		\nabla_{\alpha'}\bigg[-\frac{\sigma}{q}+\frac{(q-1)}{q}\nabla_{q}^{2}\gamma+(D-1)\gamma+\nabla_{p}^{2}\zeta\bigg]=0
	\end{gathered}
\end{equation}

We can then compute the action from the transverse traceless part of the action    
\begin{equation}
	\begin{gathered} \la{ActionTT}
		I_{TT}=\frac{1}{64 \pi G_{N}}\int \bigg[\frac{p-1}{p}\chi(-\nabla_{p}^{2})\bigg(-\nabla_{p}^{2}-\frac{p}{r_{p}^{2}}\bigg)\Delta_{0}\chi+\frac{q-1}{q}\gamma(-\nabla_{q}^{2})\bigg(-\nabla_{q}^{2}-\frac{q}{r_{q}^{2}}\bigg)\Delta_{0}\gamma \\+ 2\zeta(-\nabla_{p}^{2})(-\nabla_{q}^{2})\Delta_{0}\zeta+\frac{D}{pq}\sigma\Delta_{0}\sigma
		\\
		+x^{\alpha}\bigg(-\nabla_{p}^{2}-\frac{p-1}{r_{p}^{2}}\bigg)\Delta_{1}x_{\alpha}+y^{\alpha'}\bigg(-\nabla_{q}^{2}-\frac{q-1}{r_{q}^{2}}\bigg)\Delta_{1}y_{\alpha'}+2z^{\alpha'}(-\nabla_{p}^{2})\Delta_{1}z_{\alpha'}+2z^{\alpha}(-\nabla_{q}^{2})\Delta_{1}z_{\alpha}~~~~\\	
		+\psi^{\alpha' \beta'}\bigg(-\nabla^{2}\psi_{\alpha' \beta'}-2\tensor{R}{_{\alpha'}^{\gamma'}_{\beta'}^{\delta'}}\psi_{\gamma'\delta'}\bigg)+\psi^{\alpha \beta}\bigg(-\nabla^{2}+\frac{2}{r_{p}^{2}}\bigg)\psi_{\alpha \beta}+2\psi^{\alpha \beta'}(-\nabla^{2})\psi_{\alpha \beta'}\bigg]
	\end{gathered}
\end{equation}
where we defined the scalar operator  
\begin{equation}
\label{Laplacent}
\Delta_{0}=-\nabla^{2}-2(D-1)
\end{equation}
and the operator acting on  transverse vectors
\begin{equation} \la{VectLap}
\Delta_{1}=-\nabla^{2}-(D-1)
\end{equation}

We see from the form of the action \nref{ActionTT} as well as the constraints \nref{TosC} that the scalars, vectors and tensors decouple from each other.  

From the action \nref{ActionTT} we see that only the terms coming from the scalars and the tensors $\psi_{\alpha' \beta'}$ can be negative, because the differential operators other than $\Delta_{0}$ and the one applied to $\psi_{\alpha'\beta'}$ are all non-negative\footnote{In making this statement we are using that the operator $-\nabla_p^{2} -\frac{p}{r_{p}^{2}} $ is positive acting on $\chi$ once we exclude the modes that are constant and the ones whose gradients are conformal Killing vectors.}.  One can easily prove that $\Delta_{1}$ is non-negative, see appendix \ref{CompEin}. The possible negative modes from the scalars $\sigma$, $\chi$, $\gamma$ and $\zeta$, therefore, must be negative modes of $\Delta_{0}$ \nref{Laplacent}.

Recall that  the negative modes of $\Delta_0$ are as follows. We always have a negative mode corresponding to modes constant over the entire manifold. This mode is automatically transverse, and it is excluded from $\chi$, $\gamma$ and $\zeta$ so they are composed of $\sigma$ alone and contribute as one negative mode. 

Modes with $\ell=1$ in $S^{p}$ can only be negative modes if they are constant in the second manifold, see \nref{Delta0doub}. Such modes therefore are excluded from $\gamma$ and $\zeta$ since they are constant in $M_{q}$, and from $\chi$ since their derivatives are conformal Killing vectors in $S^{p}$. Therefore, they are fluctuations of $\sigma$ alone, which imply they cannot be made transverse since there is no such solution to \nref{TosC}. Also, modes with $\ell>1$ in $S^{p}$ are positive modes.

The remaining possible negative modes are therefore the ones which have $\ell=0$ in $S^{p}$ and are not constant over $M_{q}$. According to the bound on the scalar Laplacians, they must then satisfy $-\nabla_{q}^{2} \geq \frac{q}{r_{q}^{2}}$. If they saturate the bound their gradients are conformal Killing vectors in $M_{q}$, in which case they are excluded as we discussed above. These remaining negatives modes are constant in $S^{p}$ and eigenmodes in $M_{q}$ with $2(q-1)>-r_{q}^{2}\nabla_{q}^{2}>q$. Solving the transverse condition we find they are of the form
\begin{equation}
h_{\alpha \beta}=\frac{\sigma}{p}g_{\alpha \beta}    \text{  ,  } h_{\alpha' \beta'}= \frac{1}{(D-1)(q-\hat{\lambda}_{n})}\bigg(\nabla_{\alpha'}\nabla_{\beta'}\sigma-g_{\alpha'\beta'}\frac{[q-1-\hat{\lambda}_{n}]}{r_{q}^{2}}\sigma\bigg)
\end{equation}
where $\sigma$ is constant over $S^{p}$ and $-r_{q}^{2}\nabla_{q}^{2}\sigma=\hat{\lambda}_{n}\sigma$. As discussed in \nref{dSpMqLorAlt} these modes correspond to the $F_{n}$ modes in \cite{Hinterbichler:2013kwa}, or longitudinal negative modes as we define them, and there are ${\cal N}_{L,q}$ of them. This concludes the analysis of negative modes coming from the scalars $\sigma$ and the longitudinal fluctuations of the metric.

The only possible remaining negative modes are from $\psi_{\alpha' \beta'}$ since the rest of the action is non-negative. We do not solve explicitly for them since they depend on the details of the manifold $M_{q}$, instead we just say there are ${\cal N}_{TT,q}$ such negative eigenmodes. However, they are all scalar fields from the perspective of the $S^{p}$, or $dS_{p}$, so the QFT argument in \nref{SpScalar} implies this number matches with the number of physical negative modes obtained from transverse traceless tensors in $M_{q}$, ${\cal N}_{TT,q}$, discussed in section \nref{dSpMqLorAlt}.

The number of transverse traceless negative modes on $S^{p} \times M_{q}$ is therefore $1+{\cal N}_{L,q}+{\cal N}_{TT,q}$, and the overall phase of the path integral is, using \nref{phaseM}
\begin{equation}
Z(S^{p}\times M_{q})=(-i)^{1+{\cal N}_{L,q}+{\cal N}_{TT,q}-1}(\text{positive})=(-i)^{{\cal N}_{L,q}+{\cal N}_{TT,q}}(\text{positive}) ~,~~~~~~{\rm for }~~~p\geq 2
\end{equation}
which agrees with the answer we obtained from the KK analysis in \nref{phaseKK}, as it should. In particular, it allows us to extend the result \nref{phaseKK} to the case $p=2$.

Having discussed that we then compute the phase in specific examples. We illustrate the calculation for when $M_{q}=S^{q}$ and for when $M_{q}=S^{p_{2}} \times S^{p_{3}}\times... S^{p_{n}}$ with $n>2$. 

\subsection{$S^p \times S^q$}
\la{TwoSpheres} 

When $M_{q}=S^{q}$ the spectrum is easy to compute. In this case, there are no transverse traceless negative modes since for $S^q$ we have 
\begin{equation}
\psi^{\alpha' \beta'}[-\nabla^{2}\psi_{\alpha'\beta'}-2\tensor{R}{_{\alpha'}^{\gamma'}_{\beta'}^{\delta'}}\psi_{\gamma'\delta'}]=\psi^{\alpha'\beta'}\bigg[-\nabla^{2}+\frac{2}{r_{q}^{2}}\bigg]\psi_{\alpha'\beta'}>0
\end{equation}
therefore, ${\cal N}_{TT,q}=0$. Also, ${\cal N}_{L,q}=0$ since there are no scalar eigenvalues of $-r_{q}^{2}\nabla_{q}^{2}$ in the sphere between $q$ and $2(q-1)$, since the spectrum $\ell(\ell+q-1)$ at $\ell=1$ is exactly at $q$ and the one at $\ell=2$ is already at $2(q+1)$. Therefore, there is only one negative mode of $S^{p} \times S^{q}$ which is transverse and traceless in the full manifold, which is the constant $\phi_0$ mode \nref{VolModF}. Then we find 
\begin{equation} \la{ResSpSq}
Z(S^{p}\times S^{q})=
(\text{positive}) ~,~~~~~{\rm for }~~~~~p, q \geq 2
\end{equation}
In conclusion, there is no overall phase for the path integral in $S^{p} \times S^{q}$. The answer is symmetric under exchange of $S^{p}$ and $S^{q}$ as it should, and in a sense this is consistent with the fact that the number of physical negative modes \nref{CalNd} in section \nref{dSpMqLorAlt} is not symmetric under the exchange. From the Kaluza Klein point of view the overall phase comes from physical negative modes and the phase of the path integral $S^{p}$, so the number of physical negative modes must depend on $p$ for this combination to cancel out in $S^{p} \times S^{q}$. Also, note that we have no subtleties here when $p$ and/or $q$ are equal to two, but further subtleties arise in the next subsection where we study the product over more than two spheres.

\subsection{  $ S^{p_1} \times S^{p_{2}}\times S^{p_{3}} \times ... \times S^{p_{n}}$}
\label{ManyS}

We compute the phase of the path integral for this product of spheres using the results for a general manifold in section \ref{PhaseGen}. In this case,  there are no conformal Killing vectors on the whole manifold and all we have to do is to compute the number of negative modes of the transverse traceless tensors, ${\cal N}_{TT}$, on the whole manifold and use \nref{phaseM}.  
 
Our strategy  is to decompose $\psi_{ab}$ further in each sphere into tensors which are traceless on the individual spheres and subtraces $\phi_{i}$ such that
\begin{equation}
	\psi_{\alpha_{i}\beta_{i}}=\hat{\psi}_{\alpha_{i}\beta_{i}}+g_{\alpha_{i}\beta_{i}}\phi_{i}
\end{equation}
with $g^{\alpha_{i}\beta_{i}}\hat{\psi}_{\alpha_{i} \beta_{i}}=0$ and $\sum_{i}p_{i}\phi_{i}=0$, and $\alpha_{i}$ are the indices in sphere $S^{p_{i}}$, with $1 \leq i \leq n$. We assume that $n>2$ since we already discussed $n=2$ in the subsection above. The terms involving different spheres $\psi_{\alpha_i \beta_j}$ do not require any further decomposition. The transversity condition therefore requires that
\begin{equation}
\label{transvcond}
	\nabla_{\alpha_{i}}\phi_{i}+\nabla^{\beta_{i}}\hat{\psi}_{\alpha_{i}\beta_{i}}+\sum_{j \neq i} \nabla^{\beta_{j}}\psi_{\alpha_{i}\beta_{j}}=0
\end{equation}

Using this decomposition one can find the action for the transverse and traceless modes in $S^{p} \times M_{q}$ to be
\begin{equation}
	I_{TT}=\frac{1}{64 \pi G_{N}}\int \bigg[\sum_{i=1}^{n} \bigg[p_{i}\phi_{i}\Delta_{0}\phi_{i}+\hat{\psi}^{\alpha_{i}\beta_{i}}\bigg(-\nabla^{2}+\frac{2}{r_{p_{i}}^{2}}\bigg)\hat{\psi}_{\alpha_{i}\beta_{i}}\bigg]+2 \sum_{1=i<j}^{n}\psi^{\alpha_{i}\beta_{j}}(-\nabla^{2})\psi_{\alpha_{i}\beta_{j}}\bigg]
\end{equation}
with $\Delta_0$ in \nref{Laplacent}. 

We see that only the $\phi_i$ fields can give rise to negative modes since the other terms are positive. 
The negative modes of $\Delta_{0}$ are non-constant in at most one of the $n$ spheres, see \nref{Delta0doub}. This means that   these $\phi_{i}$ modes will either have $\ell=0$ in all spheres, or have $\ell=1$ in one sphere and $\ell=0$ in all the others. 
In principle, the condition \nref{transvcond} can lead to a mixing of $\phi_i$ and the other modes. However, we can argue that this particular form of the negative modes of $\Delta_0$ implies that the other possible components vanish. 
This is because the transverse condition \nref{transvcond} for these modes implies that $\nabla_{\alpha_{i}}\phi_{i}=0$. To see this we can integrate \nref{transvcond} against $\nabla^{\alpha_{i}}\phi_{i}$ to find, for each $i$, 
\begin{equation}
	\begin{gathered}
		0=\int[\nabla_{\alpha_{i}}\phi_{i} \nabla^{\alpha_{i}}\phi_{i}-\hat{\psi}_{\alpha_{i}\beta_{j}}\nabla^{\beta_{j}}\nabla^{\alpha_{i}}\phi_{i}-\sum_{j \neq i}\psi_{\alpha_{i}\beta_{i}}\nabla^{\alpha_{i}}\nabla^{\beta_{j}}\phi_{i}]=\int[\nabla_{\alpha_{i}}\phi_{i} \nabla^{\alpha_{i}}\phi_{i}]\\
	\end{gathered}
\end{equation}
where we integrated the last two terms by part to obtain this form. We have then set them to zero by noticing that, for the negative modes, $\phi_{i}$ has $\ell=0$ or $\ell=1$ in any sphere, so either $\nabla^{\beta_{i}}\nabla^{\alpha_{i}}\phi_{i} \sim 0$, if $\ell=0$ in that sphere, or $\nabla^{\beta_{i}}\nabla^{\alpha_{i}}\phi_{i} \sim g^{\alpha_{i}\beta_{i}}$, if $\ell=1$, in which case we use that $\hat{\phi}_{\alpha_{i}\beta_{i}}$ is traceless to set that term to zero. Also, since the mode has $\ell \neq 0$ in at most one sphere, we have that $\nabla^{\alpha_{i}}\nabla^{\beta_{j}}\phi_{i}=0$ for $i \neq j$. We are then left with a right hand side which is a sum of squares, and the only way for that to be zero is if $\nabla_{\alpha_{i}}\phi_{i}=0$. Therefore, the transverse traceless negative modes of the action are composed purely of subtrace modes, that is, the $\phi_{i}$.

It is now convenient to separate the discussion in two different cases. The first case is when all the spheres respect $p_{i}>2$, and the second case is when at least one of the spheres is two dimensional. The second case is qualitatively different because it contains physical zero modes, whose treatment goes beyond our linearized analysis. We first assume that $p_{i}>2$ for all spheres. 

\subsubsection{No $S^2$ factor, or $p_i > 2$}

Let us then discuss the possible negative modes first when all $p_{i}>2$. 

First we discuss the negative modes of $\phi_{i}$ which are constant eigenmodes over the entire manifold. They are automatically transverse, so they are only constrained by the traceless relation. Since there are $n$ different $\phi_{i}$ and 1 constraint, these are $n-1$  negative modes in ${\cal N}_{TT}$ corresponding to constant $\phi_{i}$ modes. 

Then, the only other possible negative modes are the ones with $\ell=1$ in one of the spheres and $\ell=0$ in all others. For the modes with $\ell=1$, in say the $m$-th sphere in the product, there is just  one non-trivial transversality condition involving the $\phi_{i}$ since they have by assumption $\ell=0$ in the other spheres. Since these modes, as we argued, are composed purely of subtraces $\phi_{i}$, the transverse condition reads $\nabla_{\alpha_{m}}\phi_{m}=0$, which implies that $\phi_{m}=0$ since by assumption $\phi_{m}$ has $\ell=1$ on the $m$-th sphere.

We are then only further constrained by the condition that the mode is traceless, which imply that $\sum_{i}p_{i}\phi_{i}=0$. Since we have $n-1$ variables, as $\phi_{m}$ was set to zero, and 1 constraint, there are $n-2$ independent transverse traceless linear combinations $\phi_{i}$ for a given $\ell=1$ eigenmode in one of the spheres.

For a given traceless transverse combination of the $\phi_{i}$ there are $p_{m}+1$ different modes with $l_{m}=1$ in that sphere, which is the degeneracy of these eigenmodes. Thus, if $p_{i}>2$ holds for all of the spheres, the number of ${\cal N}_{TT}$ negative modes which are non-constant over the manifold is $(n-2)\sum_{i=1}^{n}(p_{i}+1)=(n-2)(D+n)$. The number of transverse traceless negative modes in the full manifold, ${\cal N}_{TT}$, is therefore
\begin{equation}
{\cal N}_{TT}=(n-1)+(n-2)(D+n) ~,~~~~~~~\text{ for~~~  $p_{i}>2$ ~~for all spheres}
\end{equation}

This predicts from \nref{phaseM} the phase of the partition function in the full manifold to be
\begin{equation}
	\label{phasemany}
	Z[S^{p_{1}}\times S^{p_{2}}\times .... \times S^{p_{n}}]=i (-i)^{{\cal N}_{TT} }(\text{positive})=(-i)^{(n-2)(D+n+1)}(\text{positive}) ~,~~~~{\rm for } ~~~p_i > 2
\end{equation}
 Note that the answer is independent of the particular values of $p_i$ and only depends on the total number of spheres, and the total dimension $D$. 
 
\subsubsection{At least one factor of $S^{2}$}

If any of the spheres has $p_{i}=2$, then some of the negative modes we discussed are actually zero modes. This is because if $\ell=1$ in any of $S^{2}$ factors and $\ell=0$ in the others then $\Delta_{0}$ is
\begin{equation}
	\Delta_{0}=(D-1)\frac{1(1+1)}{2-1}-2(D-1)=0
\end{equation}
Note that we are assuming that $n> 2$. In the case of $n=2$ treated in section \ref{TwoSpheres} there are no physical zero modes, since we noted above that there are only $n-2$ physical fields of this kind for a given $\ell=1$ eigenmode. 

We should then discuss what we expect from the integral over these zero modes. They are not pure gauge and neither is any linear combination of them, since they are transverse and traceless, so they are physical zero modes. In principle, we need to expand the action to higher orders to understand whether they remain zero modes or they get some non-zero action at cubic or higher orders. This means that the expansion to quadratic order that we have been doing is not enough to perform the path integral.

We call the path integral result for the integral over these zero modes of the linearized action $Z_{0}$. In principle $Z_{0}$ could have a phase as well if its non-linear action has negative directions, so we do not claim $Z_{0}$ is positive. 

However, we still have a phase from the linearized fluctuations coming from the negative modes, so we discuss how the overall phase from them changes when $S^{2}$ factors are present. Since the zero modes come from eigenmodes with $\ell=1$ on the $S^{2}$ factors, we should subtract from the counting we done before for any eigenmode of this form. Each such modes was counted as $2+1=3$ negative modes, and we must subtract this overcount wherever appropriate. To correct the counting we therefore only need to change the number of negative modes coming from modes with $\ell=1$ in one of the spheres to be $(n-2)(D+n-3n_{2})$, with $n_{2}$ the number of $S^{2}$ factors in the manifold. Putting it all together we find that in general
\begin{equation}
	{\cal N}_{TT}=n-1+(n-2)(D+n-3 n_{2})
\end{equation}

The final answer for the phase of the path integral is therefore
\begin{equation}
	Z[S^{p_{1}}\times S^{p_{2}} \times... \times S^{p_{n}}]=(-i)^{(n-2)(D+n+1-3n_{2})} \times Z_{0} \times (\text{positive})
\end{equation}
and to determine the phase when $S^{2}$ factors are present, one must find the phase of $Z_{0}$, the path integral over the zero modes.

The negative modes of product of spheres in pure gravity was discussed before in \cite{Yasuda:1984py}, but we believe the author missed some negative modes.  In \cite{Yasuda:1984py}, they included the $n-1$ negative modes  with $\ell=0$ in all spheres, but missed the ones that have $\ell =1$ on one of the spheres. This difference matters only for $n>2$.

        \section{More details on the $S^2 \times M_q$ case }
        \la{TwoMq}

        \subsection{The volume field for the $p>2$ case } 
        Before we get into the details of this special case it is useful to go back to the $S^p \times M_q$ case, for $p> 2$, and give a simple description of the origin of the negative mass squared field associated to the volume of $M_q$. This volume field has a potential with two terms, one positive and one negative. The positive contribution one comes from the $D$ dimensional cosmological constant. The negative contribution comes from the positive curvature of the internal manifold.  The internal manifold has positive scalar curvature due to the Einstein equations. For small volumes this negative contribution dominates and the full potential has a positive maximum, which explains why the fluctuations around the maximum have negative mass squared. 
        More explicitly, if we denote the overall volume of $M_q$ as $v_q$, then, after going to Einstein frame on $S^p$, the potential has the form  
       \be 
       \label{potphi0}
       V \propto (v_q)^{ - {2  \over p-2 }} \left[ 1 - { C \over (v_q)^{2\over q } } \right] ~,~~~~~{\rm with}~~C> 0 ~,~~~{\rm for }~~~~~p>2
       \ee 
       which shows that it has a positive maximum. Fluctuations around this maximum lead to the mode   called $\phi_0$ around \nref{VolModF} which has a negative mass squared   \nref{MassVolM}.

        \subsection{The volume field for the $S^2 \times M_q$  case and two dimensional dilaton gravity } 
      
         In the two dimensional case we obtain a term in the action involving 
         \be \la{TwoDIm}
          - \int \sqrt{ g^{\rm or}} v_q   R^{\rm or}   + \cdots 
        \ee 
        where $g^{\rm or}_{ \mu \nu }$ denotes the original $D$ dimensional metric reduced to the two sphere. 
        Now we cannot remove the $v_q$ factor by a rescaling of the metric. 
        It is convenient, however, to perform a rescaling of the metric so as to define a new two dimensional metric $g_{\mu \nu } =(v_q)^{1-1/q} g^{\rm or}_{\mu\nu} $. This removes the kinetic term for the field $v_q$ which is the dots in \nref{TwoDIm}. For more details see Appendix \nref{RedDil}. After this, we get a final two dimensional action of the form 
        \be \la{ActVt}
        I_{E} \propto  - \int \sqrt{g} ( v_q R  - \tilde V) ~,~~~~~~~ \tilde V \propto  v_q   v_q^{-1 +1/q} \left(1 - { A \over v_q^{2/q} } \right) = v_q^{1/q} - { A \over v_q^{1/q} }   
        ~,~~~~~{\rm with } ~~~~ A> 0 
        \ee
        The $S^2 \times M_q$ background corresponds to a particular volume $v^0_q$ which is the one that sets to zero the potential $\tilde V(v^0_q) = 0$. This comes from the equation of motion for the overall scale size of the $S^2$, this mode is only present in the second term of \nref{ActVt}. We can then expand around this value as 
        \be \la{VolInt}
        v_q = v_{q,0} + \Phi 
        \ee 
  This then gives the action 
        \be   \la{JTAct}
        I_{E} = I_E^0  - \int (\Phi R - V(\Phi) ) ~,~~~~~V(\Phi) = { 2 \over r_2^2}  \Phi - C \Phi^2    ~,~~~~~~{\rm with } ~~~~C> 0 
        \ee 
       where $I_E^0$ is the action evaluated at $v^0_q$.

        It is interesting that, in this case, $\Phi$ does not contain a propagating degree  of freedom, as opposed to the case of the volume field for $p> 2$. 
        
       It is very important to note that \nref{JTAct} has a quadratic terms in $\Phi$. One might be tempted to drop this term since one would say that $\Phi$ is small. 
       
         Temporarily, {\it and incorrectly},  dropping this quadratic term we get the de Sitter JT gravity action 
       \be 
       I_{E} -I_E^0 \propto - \int \Phi (R-{ 2 \over r_2^2} ) 
       \ee  
        This de Sitter JT gravity theory has an infinite partition function \cite{Maldacena:2019cbz,Mahajan:2021nsd}. This arises because there is a family of solutions in Euclidean space with a round sphere metric and $\Phi$ given by a spin one harmonic on the $S^2$
        \be 
        ds^2 = r_2^2 \left[ d\theta^2 + \sin^2 \theta d\phi^2 \right]  ~,~~~~~~ \Phi  = \Phi_1 \cos \theta 
        \ee 
        where $\Phi_1$ is an arbitrary constant. Of course there are two other solutions we can get by rotating the sphere. Therefore the partition function is 
        \be \la{InfPart}
        Z_{JT}[S^2] \propto \int d^3 \vec \gamma_1 = \int d\gamma_1 \gamma_1^2 ~,~~~~~~~\Phi_1 = i \gamma_1   
        \ee 
        where the $\vec \gamma_1$ are (up to an $i$) the three coefficients of the $\ell=1$ harmonics. We are also emphasizing that the integral is performed along the imaginary direction for $\Phi_1$. This is because we have rotated the full dilaton field $\Phi$ into the imaginary direction so that its integral gives $R=2/r_2^2$.
        This result \nref{InfPart} is in contradiction with the result in  \nref{ResSpSq} stating that   the $S^2 \times S^q$ partition function is finite, even when $p=2$. 
        
        Fortunately, this apparent contradiction is solved by correctly including the quadratic term  in $\Phi$ in \nref{JTAct}.  Of course, for small $\Phi$ we can ignore the quadratic correction, but for the purposes of computing the partition function we should include it. 
                 When we now integrate over imaginary values of $\Phi_1 = i \gamma$ the integral has the form 
        \be \la{GammIn}
        Z  \propto \int d \gamma \gamma^2 \exp ( - C \gamma^2  ) \propto ({\rm positive } ) < \infty 
        \ee 
        
        In other words, the full $D$ dimensional Euclidean partition function computation can be Kaluza-Klein reduced to a computation in two dimensions, but we should consider the two dimensional gravity action \nref{JTAct} containing the extra quadratic term in $\Phi$,  which makes the two dimensional gravity contribution finite and positive. Of course, in the case of a more general internal manifold $M_q$ we could get additional phases from the extra negative mass squared fields that depend on the detailed geometry of $M_q$. There are no other negative modes when $M_q = S^q$.

     For $p> 2$ we had an interpretation of the phase in the partition function as due to the sphere part times a part that comes from growing quasinormal  modes in the static patch. But we do not have a precise justification of such a formula for the $p=2$ case.  However, let us give the following heuristic discussion. 
       We see in \nref{GammIn} or \nref{ResSpSq} that the partition function has no phase when $M_{q}=S^{q}$.
        If we wrote this as 
         \be \la{NoPhase}
          1 = i^{ 2 + 2} (-i)^4 
          \ee 
          we could interpret the first factor as the continuation to $p=2$ of the general pure gravity action. And one might want to argue that we could get rid of this factor by suitably including an observer as in \cite{Maldacena:2024spf}. Then the second factor in \nref{NoPhase} should be interpreted as coming from physical instabilities. As we mentioned before, the field $\Phi$ has no dynamical modes, but it does lead to three solutions, one growing like $e^t$ and one independent of time and one decreasing in time that we discard. These are precisely borderline cases in terms of Euclidean negative modes. Indeed,   we have said that in Euclidean space they are zero modes in the JT gravity approximation. If we were to count them as negative modes, by making them slightly more unstable than they actually are, then we would get precisely $4 = 1 + 3 $ from these modes and the formula 
          \nref{Ndef}. But we do not have any further justification for doing this other than it gives a nice answer.

        \subsection{Connection between the partition function computation and the norm of the Hartle Hawking wavefunction}

        It is interesting that this computation is also intimately related to the computation of the Hartle Hawking wavefunction of the universe  for a spatial slice $S^1_L \times M_q$ for a  very large circle of length $L$, see \cite{Maldacena:2019cbz,Cotler:2024xzz,Turiaci:2025xwi}.
        
        For that purpose, it is useful to start from the Euclidean and Lorentzian solutions 
        \bea 
        ds^2 = r_2^2[ d\theta^2 + \cos^2 \theta d\phi^2 ]  ~~~&\to & ~~~~ ds^2 =r_2^2[ - dt^2 + \cosh^2 t d\phi^2 ] 
        \cr 
        \Phi   = i \gamma \sin \theta ~~~~&\to & ~~~~ \Phi  = \gamma \sinh t  \la{SolCont}
        \eea
        We are then interested in evaluating the wavefunction at the time where $\Phi = \Phi_b  $, where $\Phi_b$ is a large value which we keep fixed as we consider circles of various lengths. Looking at the solution \nref{SolCont}, which is the solution that has a smooth continuation to Euclidean signature, we find that the length of the circle at $\Phi_b$ is of order 
        \be 
        L \propto e^{t_b } \propto { \Phi_b \over \gamma } 
        \ee 
        
        Now, in the computation of the partition function we were integrating over $\gamma$ in 
        \nref{SolCont}, see \nref{GammIn}. And we can express that integral as an integral over $L$ (keeping $\Phi_b$ fixed). This resulting integral has precisely the form derived from the wavefunction of the universe 
         \be 
        \int { d L \over L } |\Psi( S^1_L)   |^2 \propto \int { d L \over L} { 1 \over L^3 }  \exp( - C/L^2 )  \propto \int d \gamma \gamma^2 \exp( - C \gamma^2 ) 
        \ee 
         This formula was discussed in  \cite{Maldacena:2019cbz} where it    involved an interesting balance between one loop and classical terms.  Namely, there was factor of $1/L^{3/2}$ in $\Psi(S^1_L)$ coming   from the quantum treatment of the Schwarzian modes. The factor of $1/L$ comes from  the translation zero mode along the $S^1$  \cite{Cotler:2024xzz,Maldacena:2019cbz}. Interestingly, in the Euclidean partition function everything arises simply from a gaussian integral \nref{GammIn}. In fact, in the full computation leading to \nref{ResSpSq}  we did not even need to treat it separately from the other modes.   
        Note also that the relative $i$ between $\gamma$ and the $\Phi_1$ of the previous subsection, \nref{InfPart},  arises here from demanding that the solution \nref{SolCont} is real in Lorentzian signature.

     As a side remark, let us mention that in writing $|\Psi(S^1_L)|^2$ we are imagining we are already integrating all other fields except for the length degree of freedom. In particular, if the internal manifold $M_q$ has some isometries, as we have for $S^q$, then we are also integrating out the corresponding Kaluza-Klein gauge fields. It is clear from the definition of the Hartle Hawking wavefunction that once we integrate them out we get the partition function on the $S^2$ for those fields, which is $L$ independent\footnote{In particular, the factors of $1/L^{{\rm dim}(G)/2}$, with $G$ the isometry group of $M_q$, that we get from the analog of the Schwarzian for gauge fields,  will cancel after we integrate over the boundary values of the gauge fields   and we sum over all possible classical solutions. We thank J. Turiaci for a discussion on this topic.}.

 	 	\section{Discussion} 
 	 	
 	 	We have found that the phase of the partition function for $dS_{p}\times M_{q}$ has the form expected for pure $dS_p$, $i^{p+2}$, times extra factors that are associated to physical negative modes in the static patch. This means that interpreting the phase factor in the general $S^p \times M_q$ case is essentially the same problem as interpreting it in the case of $S^p$. A possible interpretation was suggested in \cite{Maldacena:2024spf} by introducing an observer. 
 	 	We can say something similar here, where we introduce an observer moving along the $S^p$ and in a zero momentum wavefunction on $M_q$. 
 	 	
 	 	The case of $S^p \times S^q$ is particularly interesting because in that case the total phase vanishes. This vanishing for the phase is a result of a cancellation between the contribution from a negative mass squared field and the de Sitter contribution. When we interpret this from the point of view of the static patch observer in $dS_p$, the contributions of the negative mass squared field can be interpreted as follows. We get $(-i)^{p-1}$ from the $\ell =1$ modes under the $SO(p-1)$ rotations around the center of the static patch. 	And we get $(-i)^3$ from the $\ell=0$ mode which is a single growing modes but with a growing rate $\kappa$ that is large enough to give three factors of $(-i)$.
 	 	
 	 	We can wonder whether this cancellation is in any way related to the discussion in \cite{Maldacena:2024spf} including the observer. 
 	  The $\ell =1$ modes are reminiscent of the position of the observer. On the other hand, the origin of the factor $(-i)^3$ seems different from the discussion in \cite{Maldacena:2024spf}. Here it is associated to an instability in the static patch. Perhaps it can be interpreted as an instability of the clock system for the observer in \cite{Maldacena:2024spf} which sets an origin for time. 
 	    
 	    The fact that the unstable physical field leads to a reasonable partition function with no phase seems also related to the inflationary discussion in \cite{Chen:2024rpx} where the inflaton acts like a clock.

	\subsection*{Acknowledgments}
	
	We would like to thank A. Law, G. Penington, J. Turiaci,  and E. Witten for discussions. 
	
	J.M. is supported in part by U.S. Department of Energy grant DE-SC0009988.  
	
	Z.S is supported in part by the US National Science Foundation Grant No. PHY-2209997 and by the Simons Foundation Grant No. 917464.
	
	\appendix
	
	\section{Properties of Laplacians in compact Einstein manifolds}
	\label{CompEin}
	
	Here we quickly discuss some properties of scalar and vector Laplacians in a compact Einstein Manifolds $M_{D}$ with Ricci tensor given by
	\begin{equation} \la{ECond}
	R_{a b}=\frac{(D-1)}{r_{D}^{2}}g_{a b}
	\end{equation}
	with $r_{D}$ the curvature radius of $M_{D}$. We take $r_{D}=1$ in the discussion to simplify the equations. 
		
	\subsection{Bounds on the scalar Laplacian and $\Delta_0$}
	
	\label{ScalarLap}
	
	We can obtain some spectral properties of $-\nabla^{2}$ acting on scalars by noticing that
	\begin{equation}
	\label{scalarineq}	
	 0\leq \int \big(\nabla_{a}\nabla_{b}\phi-\frac{1}{D}g_{ab}\nabla^{2}\phi\big)\big(\nabla^{a}\nabla^{b}\phi-\frac{1}{D}g^{a b}\nabla^{2}\phi\big)=\frac{(D-1)}{D}\int \phi(-\nabla^{2})(-\nabla^{2}-D)\phi
	\end{equation}
	where we noted that the left hand side quantity is non-negative, and can only be zero if $\nabla_{a}\nabla_{b}\phi=\frac{1}{D}g_{a b}\nabla^{2}\phi$ identically, which implies that $\nabla_{a}\phi$ is a conformal Killing vector. By expanding $\phi$ into a complete basis of eigenmodes of $-\nabla^{2}$ as $\phi=\sum c_{n} \phi_{n}$ with $-\nabla^{2}\phi_{n}=\lambda_{n}\phi_{n}$ we thus find that
	\begin{equation}
	\sum_{n}c_{n}^{2} \lambda_{n}(\lambda_{n}-D) \geq 0
	\end{equation}
	
	Using that the spectrum of $-\nabla^{2}$ is non-negative and that the equality above must hold for any $c_{n}$ we pick, we conclude that $\lambda_{n} \geq D$. If the inequality is saturated, it must mean that the left hand side of \nref{scalarineq} is zero and therefore $\nabla_{a}\phi$ is a conformal Killing vector of the manifold. 
	
	The converse is also true, so we conclude that $\lambda_{n}=D$ if and only if $\nabla_{a}\phi$ is a conformal Killing vector of $M_{D}$. In other words, for scalar eigenvectors of $-\nabla^{2}$, either $\lambda_{n}=0$ or $\lambda_{n} \geq D$, where the latter bound is saturated if and only if the gradients of these modes are conformal Killing vectors. For completeness, restoring the curvature radius $r_{D}$ of the manifold the bound for the excited modes is instead $\lambda_{n} \geq \frac{D}{r_{D}^{2}}$.  
		
	One can actually show that the existence of conformal Killing vectors in an Einstein manifold implies that the manifold is a sphere \cite{Obata1962CertainCF}. These conformal Killing vectors are of the form $\nabla_{a}\chi$, with $\chi$ a $\ell=1$ mode in the sphere, as we can readily check from the spectrum $-\nabla^{2}=\ell(\ell+D-1)=D$. 
	
	In the context of product manifolds the bounds can be made a bit stronger. Namely if the manifold $M_{D}$ is a product manifold of type $M_{D}=M_{p_{1}} \times M_{p_{2}} \times ... \times M_{p_{n}}$, then \nref{ECond} implies that   the curvature radii of each factor are
	\begin{equation}
	\label{radiapp}
		r_{p_{i}}^{2}=\frac{(p_{i}-1)}{(D-1)} ~,~~~~~~{\rm or }~~~~~~ R_{p_i, \, ab}  = { (p_i-1) \over r_{p_i}^2 } g_{ab} 
	\end{equation}
	which implies that the bound for the excited values for the scalars Laplacians in $M_{p_{i}}$
	\begin{equation}
		\label{boundnabp}
		-\nabla_{p_{i}}^{2}|_{\text{excited}} \geq \frac{p_{i}}{r_{p_{i}}^{2}}=\frac{p_{i}(D-1)}{p_{i}-1}
	\end{equation}
	
	Since the Laplacian in each manifold is positive, this can give a bigger lower bound for the gap of $-\nabla^{2}$, since $\frac{p}{p-1}$ is a decreasing function of $p$, and at $p=D$ we recover the previous bound. 
	
	The operator $\Delta_{0}$ defined on scalars as
	\begin{equation}
		\Delta_{0}=-\nabla^{2}-2(D-1)
	\end{equation}
	appears in the main discussion multiple times, so it is relevant to list some of its properties in the light of what was discussed in this section. Since the scalar Laplacian $-\nabla^{2}$ always has a zero mode defined by the constant eigenvalue, $\Delta_{0}$ always has a negative eigenvalue of $-2(D-1)$ corresponding to constant functions. Its first excited eigenvalue might also be negative, because the bound we have for the gap $-\nabla^{2}$ is generally less than $2(D-1)$ for $D>2$, that is
	\begin{equation}
		\Delta_{0}|_{\text{excited}}=-\nabla^{2}-2(D-1) \geq D-2(D-1)=2-D
	\end{equation} 
	
	For $D=2$, the bound guarantees that the first excited value of $\Delta_{0}$ is non-negative. In product manifolds we can also guarantee some properties of the negative modes of $\Delta_{0}$. For example, we have that in a product manifold of type $M_{p_{1}} \times M_{p_{2}} \times ... M_{p_{n}}$ the Laplacian in the full manifold is a direct sum of the Laplacians in each manifold. Also, the curvature radius of say manifold $M_{p_{i}}$, satisfies \nref{radiapp}, which implies that if the eigenfunction of $-\nabla^{2}$ are excited in more than one sphere, say on sphere $i$ and $j$, then, using \nref{boundnabp}
	\begin{equation}
	\label{Delta0doub}
		\Delta_{0}=-\nabla_{p_{1}}^{2}-\nabla_{p_{2}}^{2}-...-\nabla_{p_{n}}^{2}-2(D-1) \geq \frac{p_{i}(D-1)}{p_{i}-1}+\frac{p_{j}(D-1)}{p_{j}-1}-2(D-1)>0
	\end{equation}
	which is positive since $p/(p-1)>1$ for $p>1$. Therefore, the negative modes of $\Delta_{0}$ will be eigenfunctions which are excited in at most one manifold in the product.
	
	\subsection{Transverse vector Laplacian and $\Delta_{1}$}
	\label{transvecdelta1}

    Another relevant bound to discuss is the bound for eigenvalues of the Laplacian for transverse vectors. That is, take a transverse vector $\xi_{a}^{T}$, then it follows that
    \begin{equation}
    0 \leq \int |2\nabla_{(a}\xi_{b)}^{T}|^{2}=2 \int \xi^{a,T}(-\nabla^{2}-(D-1))\xi_{a}^{T} ~,~~~~~~~~{\rm with }~~ ~\nabla_a \xi^{a,T}=0
    \end{equation}
    
    It is clear that the left hand side is non-negative, so the eigenvectors of $-\nabla^{2}$ in transverse vectors must be bounded below by $D-1$. The saturation can only happen if the left hand side is zero, that is, $\nabla_{a}\xi_{b}^{T}+\nabla_{b}\xi_{a}^{T}=0$. In other words, the eigenvalues of $-\nabla^{2}$ for transverse vectors are bounded below by $D-1$ and the bound is saturated if and only if $\xi_{a}^{T}$ is an isometry. 
     This also implies that the operator $\Delta_{1}$ defined in \nref{VectLap}   is non-negative.

	\subsection{Various derivative identities}
	
	Here we list some identities that we found useful for the computations discussed in the main text. 
	
	Let us define the operator 
	\begin{equation}
	\Delta_{E}h_{ab}=-\nabla^{2}h_{a b}-2 \tensor{R}{_a^c_b^d}h_{c d}
	\end{equation}
    Then we find
    \begin{equation}
    \begin{gathered}
    \Delta_{E}\bigg[\bigg(\nabla_{a}\nabla_{b}-c_{1}g_{ab}\nabla^{2}\bigg)\chi\bigg]=\bigg(\nabla_{a}\nabla_{b}-c_{1}g_{ab}\nabla^{2}\bigg)(-\nabla^{2}-2(D-1))\chi\\
    \Delta_{E}[2 \nabla_{(a}\xi_{b)}^{T}]=2\nabla_{(a}[-\nabla^{2}-(D-1)]\xi_{b)}^{T}\\
    \Delta_{E}[g_{a b} \phi]=g_{a b}(-\nabla^{2}-2(D-1))\phi
    \end{gathered}
    \end{equation}
    with $c_{1}$ a generic constant.
    
    Using these results one is able to compute the quadratic action straightforwardly in terms of a decomposition like \nref{decompprod} for example. To derive them and many other formulas in the main discussion we use identities such as
    \begin{equation}
    \begin{gathered}
    \nabla^{2}\nabla_{a}\chi=\nabla_{a}[(\nabla^{2}+(D-1))\chi] \text{, } ~~~~~\nabla^{a}\nabla^{2}\xi_{a}=(\nabla^{2}+(D-1))\nabla^{a}\xi_{a}\\
    \nabla^{b}\nabla_{a}\xi_{b}=\nabla_{a}(\nabla_{b}\xi^{b})+(D-1)\xi_{a} \text{ , } \nabla^{a}\nabla^{2}h_{a b}=[\nabla^{2}+(D-1)]\nabla^{a}h_{a b}-2\tensor{R}{_a^c_b^d}\nabla^{a}h_{c d}\\
    \nabla^{2}\nabla_{a}\xi_{b}=-2\tensor{R}{_a^c_b^d}\nabla_{c}\xi_{d}+\nabla_{a}[(\nabla^{2}+(D-1))\xi_{b}]
    \end{gathered}
    \end{equation}

\section{Comments on ghost determinants}
\label{GhostDet}

For completeness, we discuss the ghost determinant for various gauge fixing conditions. Let us say that the gauge fixing condition imposed is
\begin{equation}
P_{b}(h)=\nabla^{a}h_{a b}-\betagauge \nabla_{b}h
\end{equation}
The ghost action is obtained by  using how this changes under diffeomorphisms
\begin{equation}
I_{gh} \sim \int \bar{c}^{a}[(-\nabla^{2}-(D-1))c_{a}+(1-2\betagauge)\nabla_{a}(\nabla^{b}c_{b})]
\end{equation}

We do not worry about negative modes, since one is supposed to take the absolute value of the ghost determinant \cite{Polchinski:1988ua}, but zero modes are relevant since if they exist one should deform the gauge choice slightly in order to remove them. This is important to make sure the phase obtained in the path integral for example is not ambiguous.    For convenience we decompose the ghosts $c_{a}$ as longitudinal and transverse vectors as $c_{a}=c_{a}^{T}+\nabla_{a}c^{L}$ with $\nabla^{a}c_{a}^{T}=0$, and the same for the $\bar{c}$ ghosts. In terms of this decomposition the action is
\begin{equation}
\begin{gathered}
I_{gh} \sim \int \big[\bar{c}^{T,a}(-\nabla^{2}-(D-1))c_{a}^{T}
+c_{L}(-\nabla^{2})[2(1-\betagauge)(-\nabla^{2})-2(D-1)]c_{L}\big]
\end{gathered}
\end{equation}

And we are then ready to talk about the ghost zero modes. For $c_{a}^{T}$ it is clear the zero modes are only the transverse vectors satisfying $(-\nabla^{2}-(D-1))c_{a}^{T}=0$, but as we discussed in \nref{transvecdelta1} the only solutions to these equations are isometries. For the longitudinal ghost zero modes we have that they are solutions of $[-2(1-\betagauge)\nabla^{2}-2(D-1)]c_{L}=0$. In the gauge that $\betagauge=\frac{1}{2}$, they are the solutions of $\Delta_{0}c_{L}=0$, which also defines the zero modes of many of the scalars we discussed in the main text.

More explicitly, if the manifold $M$ is a product of spheres $S^{p_{1}} \times S^{p_{2}} \times .... S^{p_{n}}$, the spectrum of $\Delta_{0}$ will be
\begin{equation}
\Delta_{0}=(D-1)\bigg[\sum \frac{\ell_{i}(\ell_{i}+p_{i}-1)}{p_{i}-1}-2\bigg]
\end{equation}

As discussed before \nref{Delta0doub},  the eigenvalue of $\Delta_{0}$ is non-positive only if $\ell=0$ in all spheres or $\ell_{i}=1$ in one sphere and zero in all others. For the mode with $\ell=0$ in all spheres $\Delta_{0}$ is negative, so it is not relevant. For the modes with $\ell=1$ in one sphere, they will be negative if the dimension of that sphere is higher than $2$, and zero otherwise. Therefore the zero modes of the longitudinal ghost are the modes with $\ell=1$ in a two dimensional sphere of the product, and $\ell=0$ in the other spheres. This implies that there are two $\ell=1$ ghost zero modes, one for $c$ and the other for $\bar{c}$, that exist for every $S^{2}$ factor in the product.  These zero modes are similar in format, but not the same as the physical zero modes we discussed in Section \nref{EucPI}.

\section{Details about Dilaton gravity}
\label{RedDil}

Consider a $D=2+q$ dimensional product manifold with the following metric 
\begin{align}
ds^2 = g_{\alpha\beta} dx^\alpha dx^\beta+g_{\alpha'\beta'} dx^{\alpha'} dx^{\beta'}, \qquad g_{\alpha'\beta'} = \rho^2\bar g_{\alpha'\beta'} 
\end{align}
In this equation, $\rho$ is a function of $x^\alpha$ which controls  the volume of the second manifold $M_q$, and   $\bar g_{\alpha'\beta'} $ is an Einstein metric which is normalized to satisfy $\bar R_{\alpha'\beta'} = q(q-1)\bar g_{\alpha'\beta'}$. For example, $\bar g_{\alpha'\beta'}$ can be the standard metric of a unit $q$-sphere.  After a lengthy calculation, we find the $D$ dimensional Ricci scalar ${\cal R}$ to be  (see \cite{Grumiller:2001ea, Grumiller:2002nm} for a derivation in Lorentzian signature)
\begin{align}
{\cal R}= R +\frac{q(q-1)}{\rho^2}\left(1-(\nabla\rho)^2\right)-\frac{2q}{\rho}\nabla^2\rho~,
\end{align}
where $R$ denotes the 2d Ricci scalar.
So the $D$ dimensional Einstein-Hilbert action becomes 
\begin{align}
I_E&=-\frac{\Omega_q}{16\pi G_N} \int \sqrt{g} \,\rho^q \left( R +\frac{q(q-1)}{\rho^2}\left(1-(\nabla\rho)^2\right)-\frac{2q}{\rho}\nabla^2\rho-2\Lambda\right)\nonumber\\
&=-\frac{\Omega_q}{16\pi G_N} \int \sqrt{g} \,\left[\rho^q \left( R -2\Lambda\right)+q(q-1)\rho^{q-2}\left(1+(\nabla\rho)^2\right)\right]~,
\end{align}
where $\Omega_q \equiv \int \sqrt{\bar g}$. To obtain the more standard form of dilaton gravity, we define $v_q (x)\equiv \rho(x)^q$ which can be thought as the volume of the internal manifold $M_q$:
\begin{align}
I_E =-\frac{\Omega_q}{16\pi G_N}  \int \sqrt{g} \left[v_q (R-2\Lambda)+\frac{q-1}{q} \frac{\left(\nabla v_q\right)^2}{v_{q}} +q(q-1) v_q^{\frac{q-2}{q}}\right]~.
\end{align}
We can further get rid of the derivative term of $v_q$ by performing the Weyl transformation 
\begin{align}\label{Weyltrans}
g_{\alpha\beta}\to v_q^{\frac{1}{q}-1}g_{\alpha\beta}~,
\end{align}
which leads to 
\begin{align}\label{IEdil}
I_E =-\frac{\Omega_q}{16\pi G_N}  \int \sqrt{g} \left(v_q R -\tilde  V(v_q)\right), \quad \tilde V(v_q) = 2\Lambda v_q^{\frac{1}{q}}-q(q-1) v_q^{-\frac{1}{q}}~.
\end{align}
Varying the scalar $v_q$ in \eqref{IEdil} sets the Ricci scalar $R = \tilde V'(v_q)$ and varying the metric yields the equation of motion for $v_q$ 
\begin{align}\label{Phieq}
 (R_{\alpha\beta}+g_{\alpha\beta}\nabla^2-\nabla_\alpha\nabla_\beta)v_q = \frac{g_{\alpha\beta}}{2} \left(v_q R- \tilde V(v_q)\right)~.
\end{align}
The two equations admit the following classical solution
\begin{align}
g_{\alpha\beta}^0 dx^\alpha d x^\beta= r_2^2 (d\theta^2+\sin^2\theta d\phi^2), \quad v_q^0 = \left(\frac{q(q-1)}{2\Lambda}\right)^{\frac{q}{2}}
\end{align}
where $v_q^0$ is the zero point of $\tilde V$, i.e. $\tilde V(v^0_q) = 0$, and the radius $r_2$ is fixed by the derivative of $\tilde V$, i.e. $\frac{2}{r_2^2} = \tilde V'(v_q^0) $. From the $D$ dimensional point of view, the value $v_q^0$ is required by the Einstein equation ${\cal R}_{ab} =\frac{2\Lambda}{ q} g_{ab}$.

Let $\Phi$ be the  fluctuation of $v_q$ around the classical value $v_q^0$, i.e. $v_q= v_q^0+\Phi$. We express the action $I_E$ in terms of this fluctuation field
\begin{align}
I_E= I_E^0-\frac{\Omega_q}{16\pi G_N}  \int \sqrt{g} \left(\Phi R - \tilde V(v_q^0+\Phi)\right), \qquad  I_E^0  =  -\frac{\Omega_q v_q^0}{16\pi G_N}  \int \sqrt{g}R~.
\end{align}
For small fluctuations, keeping terms up to $\Phi^2$, we get 
\begin{align}
I_E= I_E^0-\frac{\Omega_q}{16\pi G_N}   \int \sqrt{g} (\Phi R - V(\Phi)), \qquad V(\Phi) = \frac{2}{r_2^2}\Phi + \frac{1}{2} \tilde V''(v_q^0) \Phi^2~,
\end{align}
where $ \tilde V''(v_q^0)<0$.

In the end, let's mention that by  plugging $v_q= v_q^0+\Phi$ into \eqref{Phieq} and keeping only linear terms in $\Phi$, we find  
\begin{align}
(g_{\alpha\beta}\nabla^2-\nabla_\alpha\nabla_\beta)\Phi  = -\frac{g_{\alpha\beta}}{2} \tilde V(v_1^0+\Phi)= -\frac{g_{\alpha\beta}}{r_2^2}\Phi+\cdots
\end{align}
Taking trace on both sides yields   $r_2^2 \nabla^2 \Phi = - 2\Phi$. It can be thought as the $p=2$ limit of \eqref{MassVolM}.

\section{More on prescription I and prescription II}
\label{moreon}

\subsection{Negative gauge fixing terms}

In section \nref{ContFix} we established the correct procedure for contour rotation by demanding that the overall phase is gauge fixing independent. However, there we restricted ourselves to adding only non-negative gauge fixing terms. 

In principle, we can also add negative gauge fixing terms, but to do so consistently we have to be careful about normalization factors. To be more specific, we need to make sure that inserting the ghosts and the gauge fixing term into the path integral is the same as multiplying it by $1$.

Take a vector gauge condition $P_{a}(g)$, with $g$ parameterizing the gauge group. To gauge fix the path integral, we want to insert into it some identity of the form
\begin{equation}
N(\alpha)\int D'\omega \,e^{-\frac{\alpha}{2}\int \omega_{a}\omega^{a}}\int Dg\,\delta (P_{a}(g)-\omega_{a})|\text{Det}'\Delta_{gh}|=N(\alpha)\int D'\omega \,e^{-\frac{\alpha}{2}\int \omega^{a}\omega_{a}}=1
\end{equation}
with $Dg$ an integral over the gauge group, $\text{Det}\,\Delta_{gh}$ an appropriate ghost determinant, and $N(\xi)$ a normalization factor.

The $'$ in the integral over $\omega$, e.g in $D'\omega$, means that in integrating over $\omega$ we should restrict ourselves to vectors in the span of $P^{a}$, because otherwise the gauge condition is not achievable. For the gauge conditions we are interested in, $P^{a}$ will be orthogonal to the isometries of the manifold. Furthermore, assuming the gauge condition to be good, $P^{a}$ will be orthogonal to no other mode. 

Picking a local measure for $D\omega$ that is normalized as
\begin{equation}
\int D\omega \,e^{-\frac{1}{2}\int \omega^{a}\omega_{a}}=1
\end{equation}
we have that
\begin{equation}
N(\alpha)=\prod_{n}'\alpha^{\frac{1}{2}}=\alpha^{\frac{1}{2}(\text{V}-\text{ISO})}
\end{equation}
where the product $\prod_{n}'$ is over every local vector mode in the manifold that is in the span of $P^{a}$. This is the same as every vector mode, minus the isometry modes. We formally denote the number of vector modes of the manifold to be $\text{V}$, and the number of isometries to be $\text{ISO}$.

If $\alpha>0$ we see that $N(\alpha)$ is just a positive number. However, we could wonder what happens if $\alpha<0$. To be more specific, we would want a gauge fixing term that has negative real part, but is still regularized by some $i\epsilon$ prescription.

The reason is that the original quadratic Einstein action was proportional to $(1-i\epsilon)$. Therefore, for the gauge fixing term to combine appropriately with the original action, we need it to be slightly imaginary in the same way. To do so, for negative $\alpha$, the idea is to pick $\alpha=(-1+i\epsilon)|\alpha|$. . 

Doing so, we find that, if $\alpha<0$, $N(\alpha)$ will come with some appropriate number of factors of $i$, namely
\begin{equation}
N(\alpha)\sim i^{\Theta(-{\alpha)}(\text{V}-\text{ISO})}
\end{equation}
where $\Theta$ is the Heaviside step function. That is, there is a phase if $\alpha<0$, and not otherwise.

We can then simplify this expression even further, by using that a vector in a compact manifold with positive cosmological constant can be written as a transverse vector plus a longitudinal component\footnote{For a proof of this statement, see, for example, Appendix B of \cite{Hinterbichler:2013kwa}.} as
\begin{equation}
v_{a}=u_{a}^{T}+\nabla_{a}f
\end{equation}
with $\nabla^{a}u_{a}^{T}=0$. 

Therefore, we can alternatively think of the space of all vectors on the manifold as being the sum of the space of all transverse vectors and the space all functions, with the constant function excluded since it has zero covariant derivative. Using the prescription $\text{TV}$ to mean the number of local transverse vector modes and $\text{S}$ the number of local scalar modes, we then have the relation $V=\text{TV}+\text{S}-1$. Therefore, we can rewrite the phase of $N$ as
\begin{equation}
N(\alpha)\sim i^{\Theta(-\alpha)(\text{V}-\text{ISO})}=i^{\Theta(-\alpha)(\text{TV}-\text{ISO})}i^{\Theta(-\alpha)(\text{S}-1)}
\end{equation}

\subsection{Eigenvalue formalism}

In the main text, to argue for gauge fixing invariance of prescription I, we had to diagonalize the action using a non-local field redefinition. Then, we counted negative modes of the resulting action.

However, a more standard\cite{DeWitt:1967yk,Volkov:2000ih,Shi:2025amq}  way of computing functional determinants is as follows: First, one would compute the quadratic action of perturbations around the saddle and write that action in the form
\begin{equation}
I_{E}=\frac{1}{2}(\Phi,M\Phi)
\end{equation}
with $(\cdot ,\cdot)$ a local norm on the space of fluctuations, and $M$ an appropriate hermitian operator. Then, to compute the functional determinant one would diagonalize the operator $M$. The part of the one-loop determinant coming from non-zero eigenvalues of $M$ would then be $\text{Det}' M^{-\frac{1}{2}}$, where $\text{Det}'$ stands for the functional determinant with zero modes removed.

One might wonder if the conclusions of section \nref{ContFix} would remain the same in this formalism. In this section, we will check that this is indeed the case. Also, we will see that this formalism gives us a more natural way to distinguish between prescription I and prescription II.

First, since we are dealing with pure gravity, we need to establish a local norm on the space of metric fluctuations. We consider a real local norm of the form
\begin{equation} \la{Norbch}
(h_{ab},h_{ab})=b\int (h_{ab}h^{ab}+c h^{2})
\end{equation}
where $b$ and $c$ are arbitrary real constants. Then, once we fix a local norm, we can define a unique measure following from it, up to a normalization constant. To fix the normalization, we will impose the following relation
\begin{equation}
\label{mesnorm}
1=\int D h_{ab}e^{-\frac{1}{2}(h_{ab},h_{ab})}
\end{equation}
This equation is saying that we choose an integration contour that makes sure that all integrals are right sign gaussians. For example, if $b>0$ and $c$ is sufficiently negative in \nref{Norbch}, then, by imposing \nref{mesnorm},  we are effectively saying that we integrate $h$ along the imaginary axis.

The factor of $b$, however, only affects the measure over metrics via an overall local factor that can be absorbed into counterterms. Therefore, we can set $b$ to a particular value without loss of generality. For later convenience, we use this fact to set $b=\frac{1}{32 \pi G_{N}}$. 

It is furthermore convenient to decompose the metric fluctuation further into a pure trace part and a traceless part as
\begin{equation}
h_{ab}=g_{ab}\phi+\phi_{ab}~~, \text{with } g^{ab}\phi_{ab}=0
\end{equation}
in terms of which the local norm can be written more explicitly as
\begin{equation}
\label{locnorm}
(h_{ab},h_{ab})=\frac{1}{32 \pi G_{N}}\int (\phi_{ab}\phi^{ab}+p D\phi^{2})~~, \text{with}~~ p=1+cD
\end{equation}
Here we more clearly see that the sign of $p$ determines whether $\phi$ is integrated along the real or imaginary axis. 
%

Therefore, in this language, we can then think of prescription $I$ in section \nref{ContFix} as picking the local norm \nref{locnorm} with $p>0$, and prescription II would correspond to picking the local norm with $p<0$. 

Having established all of that, we are ready to introduce the eigenvalue problem we have to solve in pure gravity with a positive cosmological constant. 


The expression for the action is quite cumbersome, but it is slightly simpler in terms of an alternative parameterization of $\alpha$ and $\eta$ defined via
\begin{equation}
\alpha=\frac{(D-2)c^{2}}{2(D-1)\tilde{\alpha}}~~,~~ \eta D-1=\frac{(D-1)\tilde{\alpha}}{c}
\end{equation}
and in terms of which the action is
\begin{equation}
\begin{gathered}
I_{E}=\frac{1}{32\pi G_{N}}\bigg[\frac{1}{2}\int \phi^{ab}(-\nabla^{2}\phi_{ab}+2\bigg(1-\frac{(D-2)c^{2}}{2(D-1)\tilde{\alpha}}\bigg)\nabla_{(a}\nabla^{c}\phi_{b)c}-2\tensor{R}{_a^c_b^d}\phi_{cd})\\
+(D-2)(c-1)\int \phi \nabla^{c}\nabla^{c}\phi_{cd}+\frac{(D-1)(D-2)}{2}\int \phi\bigg((1-\tilde{\alpha})\nabla^{2}+D\bigg)\phi\bigg]
\end{gathered}
\end{equation}

Then, we would like to write the eigenvalue equation for $h_{ab}$. To do so, we have to write the action first in terms of overlaps defined with the norm \nref{locnorm}, more explicitly, we have
\begin{equation}
\begin{gathered}
I_{E}=\frac{1}{2}( \phi^{ab},-\nabla^{2}\phi_{ab}+2\bigg(1-\frac{(D-2)c^{2}}{2(D-1)\tilde{\alpha}}\bigg)\bigg(\nabla_{(a}\nabla^{c}\phi_{b)c}-\frac{g_{ab}}{D}\nabla^{c}\nabla^{d}\phi_{cd}\bigg)-2\tensor{R}{_a^c_b^d}\phi_{cd})\\
+\frac{(D-2)(c-1)}{2p D}(\phi,\nabla^{c}\nabla^{d}\phi_{cd})+\frac{(D-2)(c-1)}{2}(\phi^{ab},\bigg[\nabla_{a}\nabla_{b}-\frac{1}{D}g_{ab}\nabla^{2}\bigg]\phi)\\+\frac{(D-1)(D-2)}{2p D}(\phi,\big[(1-\tilde{\alpha})\nabla^{2}+D\big]\phi)
\end{gathered}
\end{equation}

From this action, we can derive the eigenvalue equation to be
\begin{equation}
\begin{gathered}
\lambda \phi=\frac{(D-1)(D-2)}{pD}((1-\tilde{\alpha})\nabla^{2}+D)\phi+\frac{(c-1)(D-2)}{pD}\nabla^{a}\nabla^{b}\phi_{ab}\\
\lambda \phi_{ab}=-\nabla^{2}\phi_{ab}-2\tensor{R}{_a^c_b^d}\phi_{cd}+2\bigg(1-\frac{(D-2)c^{2}}{2(D-1)\tilde{\alpha}}\bigg)\bigg(\nabla_{(a}\nabla^{c}\phi_{b)c}-\frac{1}{D}g_{ab}\nabla^{c}\nabla^{d}\phi_{cd}\bigg)\\
+(c-1)(D-2)\bigg(\nabla_{a}\nabla_{b}-\frac{1}{D}g_{ab}\nabla^{2}\bigg)\phi
\end{gathered}
\end{equation}

We can simplify the eigenvalue problem by further decomposing the traceless part of the metric as
\begin{equation}
\phi_{ab}=\psi_{ab}+2\nabla_{(a}\xi_{b)}^{T}+\bigg(\nabla_{a}\nabla_{b}-\frac{g_{ab}}{D}\nabla^{2}\bigg)\tilde{\chi}
\end{equation}
with $\psi_{ab}$ a transverse traceless tensor, $\xi_{a}^{T}$ a transverse vector, and $\tilde{\chi}$ a scalar. This is a convenient decomposition because these terms are all orthogonal to each other under the general norm \nref{locnorm}.

In terms of these variables, the eigenvalue problem decomposes into one for $\psi_{ab}$, one for $\xi_{a}^{T}$, and a coupled one for $\phi$ and $\tilde{\chi}$. It is therefore simpler to discuss the eigenvalues of $\xi_{a}^{T}$ and $\psi_{ab}$ first.

The eigenvalue equation for $\xi_{a}^{T}$ is
\begin{equation}
\lambda \xi_{a}^{T}=\frac{(D-2)c^{2}}{2(D-1)\tilde{\alpha}}(-\nabla^{2}-(D-1))\xi_{a}^{T}
\end{equation}

The operator $(-\nabla^{2}-(D-1))$ on transverse vectors is non-negative, and zero only when acting on isometries. However, the $\xi_{a}^{T}$ modes do not exist for isometries, since they generate no metric deformation. The non-zero eigenvalues of the $\xi_{a}^{T}$ modes thus all have the same sign, which is the sign of $\tilde{\alpha}$, which is also the sign of $\alpha$. The phase coming from the transverse vector modes is therefore
\begin{equation}
\prod_{\text{TV}}\lambda^{-\frac{1}{2}} \sim (-i)^{\Theta(-\alpha)(\text{TV}-\text{ISO})}
\end{equation}
where we excluded the isometries since these metric fluctuations do not exist.

For $\psi_{ab}$ the eigenvalue equation is
\begin{equation}
\lambda \psi_{ab}=-\nabla^{2}\psi_{ab}-2\tensor{R}{_a^c_b^d}\psi_{cd}
\end{equation}
which is independent of the gauge fixing term and $p$. The number of negative modes of $\psi_{ab}$ is finite and physical, and depends on the background. We call the number of such modes ${\cal N}_{TT}$. They contribute to the phase therefore as
\begin{equation}
\prod_{\text{TT}}\lambda^{-\frac{1}{2}} \sim (-i)^{{\cal N}_{TT}}
\end{equation}
with $\text{TT}$ standing for transverse traceless tensors.

Then, we are left only with the scalars $\phi$ and $\tilde{\chi}$. The coupled eigenvalue equation for $\phi$ and $\tilde{\chi}$ is
\begin{equation}
\begin{gathered}
\label{coupeigen}
\lambda \phi=\frac{(D-1)(D-2)}{pD}((1-\tilde{\alpha})\nabla^{2}+D)\phi+\frac{(c-1)(D-2)(D-1)}{pD^{2}}(-\nabla^{2})(-\nabla^{2}-D)\tilde{\chi}\\
\lambda \tilde{\chi}=\bigg(\frac{(D-2)}{D}\nabla^{2}+\frac{(D-2)c^{2}}{\tilde{\alpha}D}(-\nabla^{2}-D)\bigg)\tilde{\chi}+(c-1)(D-2)\phi
\end{gathered}
\end{equation}

It should be clear from \nref{coupeigen} that the $\phi$ and $\tilde{\chi}$ modes should be proportional to the same eigenmode of $-\nabla^{2}$. For concreteness, we refer to the eigenmodes of $-\nabla^{2}$ as $f_{n}$, and we call their eigenvalues $\lambda_{n}$. We should mention that $\tilde{\chi}$ generates a non-trivial metric only if $\lambda_{n}>D$. Therefore, if the condition fails, the $\tilde{\chi}$ field does not exist, and the eigenvalue equation collapses to one for $\phi$ alone.

The modes that do not fit the $\lambda_{n}>D$ criteria are those with $\lambda=D$ or $\lambda=0$. The $\lambda=0$ mode is always present and is a positive mode of $\phi$. The mode with $\lambda=D$ is a pure coordinate transformation that corresponds to a conformal Killing vector. Its sign is the same as the sign of $\tilde{\alpha}$.

Having this in mind, we can discuss the eigenvalues $\lambda$ for $\lambda_{n}>D$. They are determined by the following equation
\begin{equation}
(\lambda-\lambda_{n}^{+})(\lambda-\lambda_{n}^{-})=\lambda^{2}-a_{n}\lambda+b_{n}=0
\end{equation}
with
\begin{equation}
\begin{gathered}
\label{eqnsps}
\lambda_{n}^{+}+\lambda_{n}^{-}=a_{n}=\frac{(D-2)}{\tilde{\alpha}D}(c^{2}(\lambda_{n}-D)-\tilde{\alpha}\lambda_{n})+\frac{(D-1)(D-2)}{p D}(\tilde{\alpha}\lambda_{n}-(\lambda_{n}-D))\\
\lambda_{n}^{+}\lambda_{n}^{-}=b_{n}=-\frac{(D-1)(D-2)^{2}}{\tilde{\alpha}pD^{2}}(\tilde{\alpha}\lambda_{n}-c(\lambda_{n}-D))^{2}
\end{gathered}
\end{equation}
and where we used $\lambda_{n}^{\pm}$ to refer to the two eigenvalue solutions to equation \nref{coupeigen} for a given $n$.

Using \nref{eqnsps} we can solve for $\lambda_{n}^{\pm}$, but we will instead argue their relevant properties using their product, $b_{n}$, and their sum, $a_{n}$, assuming that the $\lambda_{n}^{\pm}$ are real. In fact, we can show that $\lambda_{n}^{\pm}$ are real as long as $\tilde{\alpha}p>0$ or $p>0$. From this, we will be able to discuss how the number of negative modes depends on the various choices for $\tilde{\alpha}$ and $p$.

For example, note that if $\tilde{\alpha}p>0$, we know that one of $\lambda_{n}^{\pm}$ is positive and the other is negative, and therefore there would be one negative mode for each $f_{n}$. 

If $\tilde{\alpha}p<0$ however, the $\lambda_{n}^{\pm}$ are two positive modes or two negative modes. Their sign is the same as the sign of their sum, $a_{n}$.

If $\tilde{\alpha}<0$ and $p>0$, we see that $a_{n}$ is negative. Therefore, in this context, the eigenvalues $\lambda_{n}^{\pm}$ are both negative. 

If $\tilde{\alpha}>0$ and $p<0$, the sign of $a_{n}$ is not guaranteed to be positive or negative, and it will in fact depend non-trivially on $c^{2}$ and $\tilde{\alpha}$. Therefore, it will not be gauge fixing invariant, which is what we found in the main text \nref{ContFix} for prescription II when the gauge fixing term was positive. In fact, the solutions for $\lambda_{n}^{\pm}$ might be complex for certain choices of gauge fixing term. If the solutions are complex, $\lambda_{n}^{\pm}$ will be complex conjugate of each other, and we will rotate them based on the sign of $a_{n}$.

We are then ready to discuss the overall phase contribution from these two scalars. First, let us focus on $p>0$. 

If $p>0$ and $\tilde{\alpha}>0$, the modes of $\phi$ and $\tilde{\chi}$ with $\lambda_{n}>D$ contribute with one negative mode per eigenmode of $-\nabla^{2}$. If $p>0$ and $\tilde{\alpha}<0$, the modes of $\phi$ and $\tilde{\chi}$ with $\lambda_{n}>D$ contribute with two negative modes per eigenmode of $-\nabla^{2}$. 

Putting this result together with the fact that $\lambda_{n}=0$ is a positive mode, and that the $\lambda_{n}=D$ modes, which correspond to conformal killing vectors, will have the same sign as $\tilde{\alpha}$, we have
\begin{equation}
\begin{gathered}
\prod_{\text{scalars}}\lambda^{-\frac{1}{2}} \sim (-i)^{\Theta(-\alpha) \,{\cal N}_{CKV}}(-i)^{(1+\Theta(-\alpha))(\text{S}-{\cal N}_{CKV}-1)}=(-i)^{\text{S}-{\cal N}_{CKV}-1}(-i)^{\Theta(-\alpha)(\text{S}-1)}
\end{gathered}
\end{equation}

Therefore, if $p>0$, the overall phase of the Euclidean path integral is
\begin{equation}
N(\alpha)\prod_{\text{TV}}\lambda^{-\frac{1}{2}}\prod_{\text{TT}}\lambda^{-\frac{1}{2}}\prod_{\text{scalars}}\lambda^{-\frac{1}{2}} \sim (-i)^{\text{S}-{\cal N}_{CKV}-1}(-i)^{{\cal N}_{TT}} \sim (-i)^{{\cal N}_{TT}-{\cal N}_{CKV}-1}
\end{equation}
which is gauge fixing invariant, and matches what we found in the main text. Therefore, we see that prescription I is gauge fixing invariant for any $\alpha \neq 0$ and any $\eta$. It also does not depend on $p$ as long as $p>0$.

Then, we might also wonder about prescription II, e.g, $p<0$. For $\tilde{\alpha}>0$, the phase is gauge fixing dependent, as discussed before, so it is not very interesting. However, it could be interesting to see what happens if we restrict ourselves to $\tilde{\alpha}<0$. 

If $\tilde{\alpha}<0$, there is one negative mode for each scalar eigenmode $f_{n}$ with $\lambda_{n}>D$, and one negative mode for each mode with $\lambda_{n}=D$. Therefore, the phase from the scalars will be
\begin{equation}
\prod_{\text{scalars}}\lambda^{-\frac{1}{2}} \sim (-i)^{{\cal N}_{CKV}}(-i)^{\text{S}-{\cal N}_{CKV}-1}=(-i)^{\text{S}-1}
\end{equation}

Therefore, the overall phase for $p<0$ and $\tilde{\alpha}<0$ is
\begin{equation}
N(\alpha)\prod_{\text{TV}}\lambda^{-\frac{1}{2}}\prod_{\text{TT}}\lambda^{-\frac{1}{2}}\prod_{\text{scalars}}\lambda^{-\frac{1}{2}}\sim (-i)^{{\cal N}_{TT}} 
\end{equation}

Therefore, for $\tilde{\alpha}<0$, prescription II gives a very satisfying answer. First, the prescription seems to be gauge fixing invariant as long as we restrict ourselves to $\tilde{\alpha}<0$. Secondly, from this result, the phase of the manifold comes only from transverse traceless negative modes. This result would imply, in particular, that the sphere partition function has no phase. 

However, the fact that prescription II seems to be scheme dependent makes one doubt if its result for $\tilde{\alpha}<0$ is reliable in the first place. In other words, one should not be using a prescription that is scheme dependent.

To go around this issue, one would have to argue that, for some reason, the gauge fixing parameters should be restricted to $\tilde{\alpha}<0$, and that there is something problematic with picking $\tilde{\alpha}>0$. The existence of such an argument seems unlikely, at least for the prescription as stated, since there is a prescription, prescription I, that seems consistent for any sign of $\tilde{\alpha}$.
    
\bibliographystyle{apsrev4-1long}
\bibliography{GeneralBibliography.bib}
\end{document}